\documentclass[a4paper,11pt]{article}

\pdfoutput=1

\usepackage{amsmath,amssymb,mathtools}
\usepackage{color}
\usepackage{graphicx}
\usepackage{subfigure}
\usepackage{cite}
\usepackage[colorlinks=true,linkcolor=blue, citecolor=red, urlcolor=blue, bookmarks]{hyperref}
\usepackage{multirow,makecell}
\usepackage{textcomp}
\usepackage{wasysym}

\usepackage[text={17.1cm,24.6cm},centering]{geometry} 

\numberwithin{equation}{section}

\def \be {\begin{equation}}
\def \ee {\end{equation}}
\def \ba {\begin{array}}
\def \ea {\end{array}}
\def \bea{\begin{eqnarray}}
\def \eea{\end{eqnarray}}
\def \nn {\nonumber}

\def \a {\alpha}
\def \b {\beta}
\def \g {\gamma}

\def \G {\Gamma}
\def \d {\delta}
\def \D {\Delta}
\def \e {\epsilon}
\def \ve {\varepsilon}
\def \m {\mu}
\def \n {\nu}

\def \l {\lambda}

\def \s {\sigma}

\def \r {\rho}
\def \o {\omega}
\def \O {\Omega}
\def \th {\theta}

\def \vph {\varphi}

\def \t {\tau}

\def \cA {\mathcal A}

\def \cL {\mathcal L}

\def \cO {\mathcal O}
\def \cP {\mathcal P}

\def \cS {\mathcal S}

\def \rC {\mathrm C}

\def \rR {\mathrm R}

\def \rZ {\mathrm Z}

\def \mA {\mathcal A}

\def \p {\partial}

\def \vn {\varnothing}
\def \f {\frac}

\def \lt {\left}
\def \rt {\right}

\def \sr {\sqrt}
\def \td {\tilde}

\def \pp {\propto}
\def \inf {\infty}

\def \lag {\langle}
\def \rag {\rangle}

\def \dd {\mathrm{d}}
\def \ep {\mathrm{e}}
\def \ii {\mathrm{i}}

\def \tr {\textrm{tr}}

\def \and {{~\textrm{and}~}}

\def \CFT {{\textrm{CFT}}}
\def \SC {{\textrm{SC}}}

\def \NS {{\textrm{NS}}}

\def \va {\vec{\alpha}}
\def \vb {\vec{\beta}}
\def \vz {\vec{z}}
\def \vm {\vec{m}}

\def \Thab {\Theta\Big[{}^{\vec{\alpha}}_{\vec{\beta}}\Big]}
\def \thab {\theta\big[{}^{\alpha}_{\beta}\big]}

\begin{document}

\title{\textbf{R\'enyi entropy and subsystem distances in finite size and thermal states in critical XY chains}}
\author{
Ra\'ul Arias$^{1,2}$\footnote{rarias@sissa.it},
Jiaju Zhang$^1$\footnote{jzhang@sissa.it}
}
\date{}

\maketitle
\vspace{-10mm}
\begin{center}
{\it
$^{1}$SISSA and INFN, Via Bonomea 265, 34136 Trieste, Italy\\
\vspace{1mm}
$^{2}$Instituto de F\'isica La Plata - CONICET and Departamento de F\'isica, Universidad Nacional de La Plata
C.C. 67, 1900, La Plata, Argentina
}
\vspace{10mm}
\end{center}

\begin{abstract}
  We study the R\'enyi entropy and subsystem distances on one interval for the finite size and thermal states in the critical XY chains, focusing on the critical Ising chain and XX chain with zero transverse field. We construct numerically the reduced density matrices and calculate the von Neumann entropy, R\'enyi entropy, subsystem trace distance, Schatten two-distance, and relative entropy. As the continuum limit of the critical Ising chain and XX chain with zero field are, respectively, the two-dimensional free massless Majorana and Dirac fermion theories, which are conformal field theories, we compare the spin chain numerical results with the analytical results in CFTs and find perfect matches in the continuum limit.
\end{abstract}

\baselineskip 18pt
\thispagestyle{empty}
\newpage


\tableofcontents

\section{Introduction}

Quantum entanglement has become one of the key tools to the understanding of the quantum many-body systems and quantum field theories
\cite{Amico:2007ag,Eisert:2008ur,calabrese2009entanglement,Laflorencie:2015eck,Witten:2018lha}.
For a quantum system in a state with the density matrix $\r$, one could choose a subsystem $A$ and trace out the degrees of freedom of its complement $\bar A$ to get the reduced density matrix (RDM) $\r_A=\tr_{\bar A} \r$ of the subsystem.
With the RDM $\r_A$, one could compute the von Neumann entropy
\be
S_A = - \tr_A ( \r_A \log \r_A ),
\ee
and R\'enyi entropy
\be
S_A^{(n)} = -\f{1}{n-1} \log \tr_A \r_A^n.
\ee
The $n\to1$ limit of the R\'enyi entropy gives the von Neumann entropy
\be
S_A = \lim_{n \to 1} S_A^{(n)}.
\ee
When the whole system is in a pure state $\r=|\Psi\rag\lag\Psi|$, the von Neumann entropy is a rigorous measure of the entanglement, which is usually called the entanglement entropy,
but in cases where the whole system is in a mixed state neither the von Neumann entropy nor the R\'enyi entropy is a good entanglement measure.
Nevertheless they are still interesting quantities that characterize to some extent the amount of entanglement.

In this paper we will consider a subsystem $A$ that is an interval of length $\ell$ in a one-dimensional quantum system, and it has different RDMs $\r_A$ in different states $\r$ of the total system.
The most general case we will consider is an interval on a torus with spatial circumference $L$ and imaginary temporal period $\b$, which is a finite system in a thermal state. We denote the RDM of the interval in such a state as $\r_A(L,\b)$.
Taking $\b \to \inf$ limit we get an interval on a vertical cylinder with spatial period $L$, which is a finite system in the ground state. We denote the RDM in such a state as $\r_A(L)$.
On the other hand, taking $L \to \inf$ for the torus, we get an interval on a horizontal cylinder with imaginary temporal period $\b$, which is an infinite system in a thermal state. We denote the RDM in such a state as $\r_A(\b)$.
Taking both $L \to \inf$ and $\b \to \inf$ limit, we get an interval on a complex plane, which is an infinite system in the ground state.
We will denote the RDM in such a state as $\r_A(\vn)$.

The continuum limit of one-dimensional critical quantum spin chains could be described by two-dimensional (2D) conformal field theories (CFTs) \cite{Cardy:1984rp,Cardy:1986ie,Bloete:1986qm,Cardy:1986Logarithmic,Affleck:1986bv}.
Some examples are the continuum limit of the critical Ising chain, which is the 2D free massless Majorana fermion theory and is a 2D CFT with central charge $c=\f12$, and the continuum limit of the XX chain with zero transverse field that gives the 2D free massless Dirac fermion theory, or equivalently the 2D free massless compact boson theory with the unit radius target space, which is a 2D CFT with central charge $c=1$.
The spin chains at critical points demonstrate universal properties that are captured by the corresponding CFTs, and it is interesting to compare various quantities in critical spin chains with the CFT predictions. In this paper we will consider the von Neumann and R\'enyi entropies.
Some examples are the cases of one interval in the ground state \cite{Holzhey:1994we,Vidal:2002rm,Latorre:2003kg,Calabrese:2004eu} and excited states \cite{Alcaraz:2011tn,Berganza:2011mh,Taddia:2016dbm}, and the cases of multiple intervals in the ground state \cite{Furukawa:2008,Casini:2008wt,Facchi:2008Entanglement,Caraglio:2008pk,Calabrese:2009ez,Alba:2009ek,Igloi:2009On,Fagotti:2010yr,Calabrese:2010he,Alba:2011fu,%
Rajabpour:2011pt,Coser:2013qda,DeNobili:2015dla,Coser:2015dvp,Ruggiero:2018hyl,Arias:2018tmw}.
In this paper, we consider the case of one interval in a state with both a finite size and a finite temperature in the critical XY chains.
We focus on two special critical points of the spin-$\f12$ XY chain, i.e. the critical Ising chain and the XX chain with zero field.
In a 2D CFT, the state with both a finite size and a finite temperature is described by the theory on a torus.
To calculate the R\'enyi entropy on a torus in the 2D free massless boson and fermion theories, one needs to take into account properly the various boundary conditions and spin structures on the replicated multi-genus Riemann surface. The final complete results were given in \cite{Chen:2015cna,Mukhi:2017rex}, and previous results could be found in %
\cite{Azeyanagi:2007bj,Ogawa:2011bz,Herzog:2012bw,Herzog:2013py,Barrella:2013wja,Datta:2013hba,Cardy:2014jwa,Chen:2014unl,Lokhande:2015zma,Klich:2017qmt,%
Blanco:2019xwi,Fries:2019ozf,Fries:2019acy}.

The motivation of the paper is twofold.
The first is to check the CFT R\'enyi entropy on a torus, which is difficult to calculate and it took several years from people first considered the problem \cite{Azeyanagi:2007bj} to finally found the complete solution \cite{Chen:2015cna,Mukhi:2017rex}.
The CFT von Neumann entropy on a torus has not been worked out, and we will calculate the leading order von Neumann entropy in short interval expansion.
On the other hand, the construction of the RDMs in spin chain finite size and thermal states has not been considered, and we will elaborate on how to do it and calculate the von Neumann and R\'enyi entropies based on the numerical construction.
We will compare the analytical CFT results of the von Neumann and R\'enyi entropies and the numerical spin chain results and find perfect matches in the continuum limit.

Often knowing the entanglement is not enough to characterise the system, and it is also interesting to know quantitatively the difference between two density matrices  \cite{nielsen2010quantum,hayashi2017quantum,watrous2018theory}.  In the framework of quantum information theory there are many quantities that do this job like for example the relative entropy, fidelity, Bures distance, trace distance, Schatten distance, and the quantum relative R\'enyi entropies.
Each of them has different quantum properties and because of this, the choice of which one of them is more useful depends on the problem at hand and the difficulty to compute it.
For example, studying the relative entropy of a pair of density matrices (on top of the information about the distinguishability of the states) one can also obtain information about the modular Hamiltonian (also called entanglement Hamiltonian) of the theory (see \cite{Arias:2016nip,Arias:2017dda});
studying the fidelity one can also detect the location (in the parameter space of the theory) of phase transitions  \cite{Gu:2008Fidelity}.
As a last relevant example in high energy physics it is was shown in \cite{Suzuki:2019xdq,Kusuki:2019hcg} that measuring the Bures distance one can construct the entanglement wedge defined in the holographic dual of the CFT.

As we mentioned above, there are many objects typically studied in quantum information theory that measures the distinguishability between different states that can be useful in CFTs. In the present work we will just analyse some of them, i.e. the trace distance, the Schatten $n$-distance and the relative entropy.
For two density matrices $\r,\s$, the trace distance is defined as \cite{nielsen2010quantum,hayashi2017quantum,watrous2018theory}
\be
D(\r,\s) = \f{\tr | \r-\s|}2.
\ee
Subsystem trace distances in low-lying energy eigenstates and states after local operator quench in 2D CFTs and one-dimensional quantum spin chains have been investigated \cite{Zhang:2019wqo,Zhang:2019itb,Zhang:2019kwu}. In these works the replica trick was used
\be
\tr | \r-\s| = \lim_{n_e \to 1} \tr(\r-\s)^{n_e},
\ee
and one firstly evaluates the right hand side for a general even integer $n_e$ and then makes the analytic continuation to one $n_e \to 1$.%
\footnote{The trick is similar to the calculation of the entanglement negativity in \cite{Calabrese:2012ew,Calabrese:2012nk}.}
For $n\geq 1$, one could also define the Schatten $n$-distance
\be
D_n(\r,\s) = \f{( \tr | \r-\s|^n )^{1/n}}{2^{1/n}}.
\ee
In 2D CFT, the Schatten $n$-distance defined above for two RDMs $\r_A,\s_A$ depends on the UV cutoff, and we will add a normalization to cancel this divergence.
So, as in \cite{Zhang:2019itb} we are going to work with the following quantity
\be
D_n(\r_A,\s_A) = \Big( \f{ \tr_A | \r_A-\s_A |^n}{2 \tr_A( \r_A(\vn)^n)}\Big)^{1/n}.
\ee
Remember that $\r_A(\vn)$ is the RDM of the subsystem $A$ on an infinite system in the ground state.
Another quantity that characterizes the difference between two states $\r,\s$ is the relative entropy
\be
S(\r\|\s) = \tr (\r \log \r) - \tr (\r\log\s).
\ee
We will calculate the subsystem trace distance, the Schatten two-distance and the relative entropy among these RDMs $\r_A(L,\b)$, $\r_A(L)$, $\r_A(\b)$, $\r_A(\vn)$ in both CFTs and spin chains and compare the results.

The remaining part of the paper is arranged as follows.
In section~\ref{secIsing}, we consider the critical Ising chain and the 2D free massless Majorana fermion theory.
In section~\ref{secXX}, we consider the XX chain with zero field and the 2D free massless Dirac fermion theory.
In these two sections, we compare the CFT and spin chain results of von Neumann entropy, R\'enyi entropy, subsystem trace distance, Schatten two-distance, and relative entropy, and find perfect matches in the continuum limit.
We conclude with discussions in section~\ref{secCon}.
In appendix~\ref{appBDT}, we show that the method of twist operators cannot give the correct short interval R\'enyi entropy on a torus at the order $\ell^4$ in some specific 2D CFTs, including the 2D free massless Majorana and Dirac fermion theories.
In appendix~\ref{appRDM}, we elaborate on how to construct the numerical RDMs in the finite size and thermal states in the XY chains, especially in the critical Ising chain and the XX chain with zero field.
In appendix~\ref{appSRE} we compare the CFT and spin chain results of subsystem relative entropy among low-lying energy eigenstates.

\section{Critical Ising chain}\label{secIsing}

We consider the critical Ising chain, whose continuum limit gives a 2D free massless Majorana fermion theory, which is a 2D CFT with central charge $c=\f12$.

\subsection{von Neumann and R\'enyi entropies}

We will first review the result for the R\'enyi entropy of one interval $A=[0,\ell]$ on a torus in the 2D free massless Majorana fermion theory \cite{Mukhi:2017rex}, and then we will recompute it using twist operators \cite{Calabrese:2004eu,Cardy:2007mb,Calabrese:2009qy} and their operator product expansion (OPE) \cite{Headrick:2010zt,Calabrese:2010he,Rajabpour:2011pt,Chen:2013kpa,Lin:2016dxa,Chen:2016lbu,Ruggiero:2016khg,He:2017txy}.
We get the same R\'enyi entropy to order $\ell^2$ from OPE of twist operators as from the expansion of the exact result in \cite{Mukhi:2017rex}.
The short interval expansion of the R\'enyi entropy allows us to do the analytic continuation $n \to 1$ and obtain the von Neumann entropy to order $\ell^2$.

In the critical Ising chain, we construct numerically the RDMs in the finite size and thermal states and compute the von Neumann entropy for a short interval and the R\'enyi entropy for a relatively long interval.
We compare the analytical CFT results with the numerical data for the spin chain and find perfect matches in the continuum limit.

\subsubsection{CFT results}

Details of the 2D free massless Majorana fermion theory can be found in the books \cite{DiFrancesco:1997nk,Blumenhagen:2009zz}.
Apart from the identity operator 1 in the Neveu-Schwarz (NS) sector, there is a primary operator $\s$ with conformal weights $(\f{1}{16},\f{1}{16})$ in the Ramond (R) sector and a primary operator $\ve$ with conformal weights $(\f{1}{2},\f{1}{2})$ in the NS sector.

The state with both a finite size and a finite temperature in 2D CFT corresponds to a torus which in our case has spatial period $L$ and temporal period $\b$, the interval $A$ has length $\ell$.
The R\'enyi entropy of one interval on a torus was computed in \cite{Mukhi:2017rex} from higher genus partition function, and it was argued in \cite{Mukhi:2017rex,Mukhi:2018qub} that the method of twist operators cannot give the correct answer for a fermion theory.
The result can be written in terms of the ratio $x={\ell}/{L}$ and the torus modulus $\t={\ii\b}/{L}$.
The R\'enyi entropy of the interval $A$ on the torus is \cite{Mukhi:2017rex}
\be \label{REmmw}
S_A^{(n)} = \f{n+1}{12n}\log\Big| \f{L}{\e} \f{\th_1(x|\t)}{\th_1'(0|\t)} \Big|
           -\f{1}{n-1} \log\bigg[ \f{\sum_{\va,\vb}\Big| \Thab(0|\O) \Big|}
                                    {\big(\prod_{k=1}^{n-1}|A_k|\big)^{1/2}\big( \sum_{\n=2}^4|\th_\n(0|\t)| \big)^n } \bigg],
\ee
with the period matrix of the higher genus Riemann surface
\be \label{Odef}
\O_{ab}(x,\t) = \f{1}{n} \sum_{k=0}^{n-1} \cos\Big[ \f{2\pi(a-b)k}{n}\Big] C_k(x,\t), ~~ C_k(x,\t) = \f{B_k(x,\t)}{A_k(x,\t)},
\ee
and
\bea \label{ABodef}
&& A_k(x,\t) = \int_{\f{\t}{2}}^{1+\f{\t}{2}}\!\!\o(z,x,\t)\dd z, ~~
   B_k(x,\t) = \int_{\f{1}{2}}^{\f{1}{2}+\t}\!\!\o(z,x,\t)\dd z, \nn\\
&& \o(z,x,\t) = \f{\th_1(z|\t)}{\th_1\big(z+\f{k}{n}x|\t\big)^{1-\f{k}{n}}\th_1\big(z-(1-\f{k}{n})x|\t\big)^{\f{k}{n}}}.
\eea
In $A_k$, $B_k$, we have shifted the integral ranges to make them convenient for numerical evaluation.

The genus-$n$ Siegel theta function is defined as
\be
\Thab(\vz|\O) = \sum_{\vm\in\rZ^n} \exp\big[ \pi\ii(\vm+\va)\cdot\O\cdot(\vm+\va) + 2\pi\ii(\vm+\va)\cdot(\vz+\vb) \big],
\ee
with $\cdot$ being multiplications between vectors and matrices.
The entries of the $n$-component vectors $\va,\vb$ are chosen independently from 0 and $\f12$ and the sum of $\va,\vb$ in (\ref{REmmw}) is over all the possible spin structures.
The Jacobi theta function is
\be
\thab(z|\t) = \sum_{m\in\rZ} \exp\big[ \pi\ii\t(m+\a)^2 + 2\pi\ii(m+\a)(z+\b) \big],
\ee
and, as usual, we have the relations
\be
\theta_1(z|\t)=-\theta\big[{}^{1/2}_{1/2}\big](z|\t), ~~
\theta_2(z|\t)=\theta\big[{}^{1/2}_{~0}\big](z|\t), ~~
\theta_3(z|\t)=\theta\big[{}^{0}_{0}\big](z|\t), ~~
\theta_4(z|\t)=\theta\big[{}^{~0}_{1/2}\big](z|\t).
\ee

Following \cite{Chen:2016lbu}, we can use the OPE of twist operators to obtain the short interval expansion of the R\'enyi entropy
\be\label{REsie}
S_A^{(n)} = \f{n+1}{12n}\log\f{\ell}{\e} - \f{(n+1)\ell^2}{6n} \Big( \lag T \rag + \f14 \lag \ve \rag^2 \Big) + O(\ell^4),
\ee
where the expectation values on the torus read \cite{DiFrancesco:1997nk}
\be
\lag T \rag = - \f{2\pi^2 q}{L^2} \f{\p_q Z(q)}{Z(q)}, ~~
\lag \ve \rag = \f{\pi}{L} \f{\eta(\t)^2}{Z(q)}.
\ee
Here we set $q=\ep^{2\pi\ii\t}$ and the partition function can be written as\footnote{In this paper we only consider the case without the chemical potential, i.e. that $\t$ is purely imaginary, and so $\bar q=q$. We have the partition function $Z(q)=Z(q,\bar q=q)$, and $\lag T \rag = \lag \bar T \rag$.}
\be
Z(q) = \f{1}{2\eta(\t)}\big[ \th_2(0|\t)+\th_3(0|\t)+\th_4(0|\t) \big].
\ee

\begin{figure}
  \centering
  \includegraphics[height=0.25\textwidth]{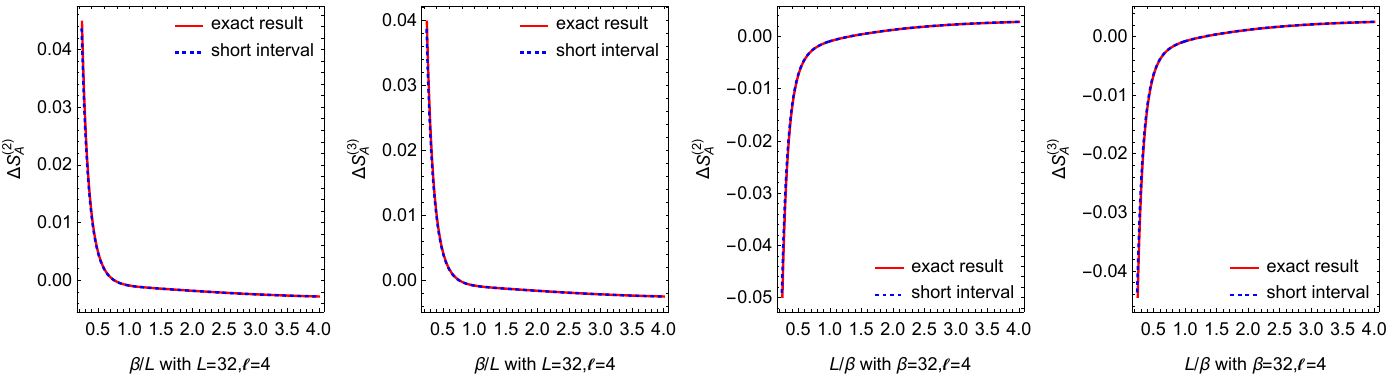}\\
  \caption{The comparison of the exact R\'enyi entropy with the short interval expansion one in the free massless Majorana fermion theory.
  We use $\D S_A^{(n)} = S_A^{(n)} - \f{n+1}{12n}\log\f{\ell}{\e}$ to make it independent of the UV cutoff.
  }\label{MajoranaRE2RE3}
\end{figure}

The short interval expansion of R\'enyi entropy (\ref{REsie}) is consistent with the small $\ell$ expansion of the exact result (\ref{REmmw}), which is
\be\label{REsie2}
S_A^{(n)} = \f{n+1}{12n}\log\f{\ell}{\e}
          + \f{(n+1)\ell^2}{24nL^2}
           \Big[ \f13 \f{\th_1'''(0|\t)}{\th_1'(0|\t)}
                -\f{\sum_{\n=2}^4\th_\n''(0|\t)}{\sum_{\n=2}^4\th_\n(0|\t)}
                -\Big(\f{\th_1'(0|\t)}{\sum_{\n=2}^4\th_\n(0|\t)}\Big)^2
           \Big] + O(\ell^{4}).
\ee
Note that $\th_1(0|\t)=\th_\n'(0|\t)=\th_1''(0|\t)=\th_\n'''(0|\t)=0$ with $\n=2,3,4$ and using the identities
\be
\th_1'(0|\t) = 2\pi \eta(\t)^3, ~~ q \p_q \th_\n(z|\t) = - \f{1}{8\pi^2} \th''_\n(z|\t), ~ \n=1,2,3,4,
\ee
we can show that the expressions (\ref{REsie}) and (\ref{REsie2}) are in fact the same.
This means that the method of short interval expansion from OPE of twist operators is valid at order $\ell^2$.
However, it breaks down at order $\ell^4$, as we show in appendix~\ref{appBDT}.
For a short interval, we compare the exact R\'enyi entropy and the short interval expansion in Fig.~\ref{MajoranaRE2RE3}.
We have subtracted the R\'enyi entropy of the same interval $A$ on an infinite straight line in the ground state to make it independent of the UV cutoff, i.e. we use
\be
\D S_A^{(n)} = S_A^{(n)} - \f{n+1}{12n}\log\f{\ell}{\e}.
\ee
We see good matches for the exact and leading order short interval results.
This is an indication that the small $\ell$ expansion for the R\'enyi entropy is a good approximation in the regime of parameters we consider.

The short interval result (\ref{REsie}) remarks the validity of the method of twist operators at the order $\ell^2$ in the small $\ell$ expansion.
Furthermore, it is convenient to do the analytic continuation $n\to1$ and get the short interval expansion of the von Neumann entropy
\be\label{EEsie}
S_A = \f{1}{6}\log\f{\ell}{\e} - \f{\ell^2}{3} \Big( \lag T \rag + \f14 \lag \ve \rag^2 \Big) + O(\ell^4).
\ee

\subsubsection{Spin chain results}

We will compare the R\'enyi entropy on a torus in the free massless Majorana fermion theory with the R\'enyi entropy for a thermal state in a periodic critical Ising chain.
In order to do that the numerical RDM of one interval in the finite size and thermal states in critical Ising chain is going to be computed following \cite{chung2001density,Vidal:2002rm,peschel2003calculation,Latorre:2003kg,Fagotti:2010yr}, as detailed in appendix~\ref{appRDM}.
To handle the zero modes in the R sector in critical XY chains, we will need a special trick as was first studied in \cite{Fagotti:2010yr}.
To compute the von Neumann entropy, we will need the explicit numerical RDMs and, unfortunately, we can only compute it for a short interval.
For the R\'enyi entropy, the correlation matrices are enough, and then we can calculate it for a relatively long interval.

\begin{figure}[htbp]
  \centering
  \includegraphics[height=0.4\textwidth]{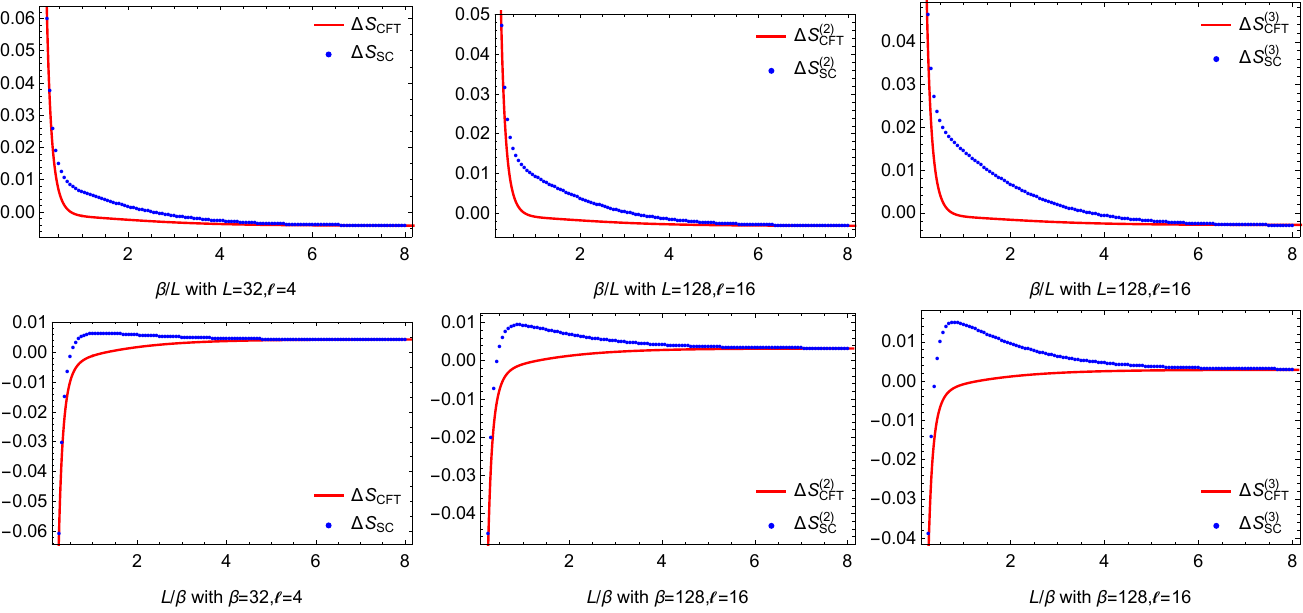}\\
  \caption{We compare the von Neumann and R\'enyi entropies in the free massless Majorana fermion theory with the numerical results in the critical Ising chain. We see deviations of the results that we attribute to finite values of $L,\b,\ell$.}
  \label{IsingEERE2RE3}
\end{figure}

On the CFT side, we use the short interval expansion of the von Neumann entropy (\ref{EEsie}) and the exact R\'enyi entropy (\ref{REmmw}). Let us start setting the nomenclature for the objects we will compute. We call the CFT von Neumann and R\'enyi entropies as $S_\CFT(L,\b)$ and $S_\CFT^{(n)}(L,\b)$ and the spin chain von Neumann and R\'enyi entropies as $S_\SC(L,\b)$ and $S_\SC^{(n)}(L,\b)$.
The CFT and spin chain results are compared in Fig.~\ref{IsingEERE2RE3}.
Note that in the CFT we have the subtracted CFT results of the von Neumann and R\'enyi entropies on an infinite line in the ground state to obtain $\D S_\CFT(L,\b)$ and $\D S_\CFT^{(n)}(L,\b)$, and in the spin chain the subtracted results of the von Neumann and R\'enyi entropies on an infinite chain in the ground state are called $\D S_\SC(L,\b)$ and $\D S_\SC^{(n)}(L,\b)$.
In other words, $\D S_\CFT(L,\b)$ and $\D S_\CFT^{(n)}(L,\b)$ are pure CFT results, $\D S_\SC(L,\b)$ and $\D S_\SC^{(n)}(L,\b)$ are pure spin chain results, and we have compared results independently obtained in CFT and spin chain.
Unfortunately, in Fig.~\ref{IsingEERE2RE3} there are generally no good matches between the analytical CFT and numerical spin chain data.
As $L \gg \b$ and $L \ll \b$, the matches are good, but for general $L,\b$, especially for $L/\b\sim1$, there are large deviations.
We believe the derivations are due to finite values of $L,\b,\ell$.

\begin{figure}[htbp]
  \centering
  \includegraphics[height=0.25\textwidth]{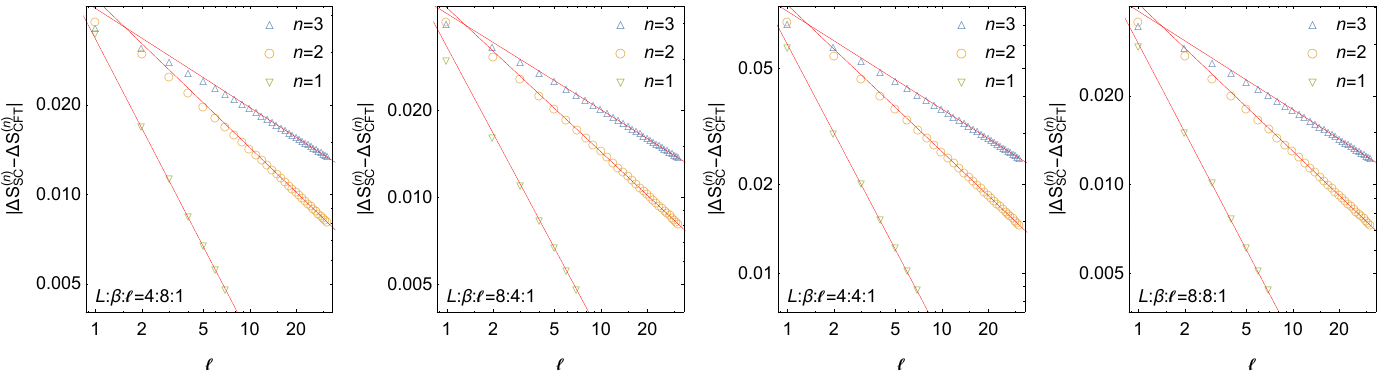}\\
  \caption{The difference of the von Neumann and R\'enyi entropy in the critical Ising chain and the free massless Majorana fermion theory with increase of $\ell$.
  We see perfect matches in the continuum limit of the spin chain.
  The thin solid red lines are proportional to $\ell^{-2/n}$.}
  \label{IsingEERE2RE3Continuum}
\end{figure}

To better see the continuum limit of the critical Ising chain, we fix the ratios $L:\b:\ell$, which make the scale invariant CFT result $\D S_\CFT^{(n)}$ a constant, and look into the difference between the von Neumann and R\'enyi entropies in spin chain and tne ones in CFT with the increase of interval length $\ell$. We plot the results in Fig.~\ref{IsingEERE2RE3Continuum}. We see that the differences of spin chain and CFT results decrease monotonically. Furthermore, by numerical fit, we get approximately
\be
| \D S_\SC^{(n)} - \D S_\CFT^{(n)} | \pp \ell^{-1/n}.
\ee
Thus we obtain perfect matches between the CFT and spin chain results of the von Neumann and R\'enyi entropies in the continuum limit of the spin chain.

\subsection{Trace distance}

We consider the short interval expansion of the subsystem trace distance.
The leading order trace distance of two RDMs $\r_A,\s_A$ depends on the quasiprimary operators with the lowest scaling dimension that have different expectation value in the two states $\r,\s$.
Among the states on the plane and cylinders $\r(\vn)$, $\r(L)$, and $\r(\b)$, the quasiprimary operators that satisfy these properties are the stress tensor $T$, $\bar T$.
Furthermore, they always have the same expectation values $\lag T \rag_\r=\lag \bar T \rag_\r$ in one of such states $\r(\vn)$, $\r(L)$, and $\r(\b)$ (that we denote here by $\r$) and 
\be \label{DrAsAMajorana2}
\lag T \rag_\r - \lag T \rag_\s = \lag \bar T \rag_\r - \lag \bar T \rag_\s.
\ee
Following \cite{Zhang:2019wqo,Zhang:2019itb}, we can use OPE of twist operators to get the leading order of the short interval expansion for the trace distance
\be \label{TheFormula}
D(\r_A,\s_A) = \f{y_T\ell^2}{\sr{2 c}} | \lag T \rag_\r - \lag T \rag_\s | + o(\ell^2).
\ee
We have the coefficient
\be \label{TheCoefficient}
y_T = \lim_{p\to\f12} \Big(\f{2}{c}\Big)^p \sum_{\cS \subseteq \cS_0}
\Big[ \Big\lag \prod_{j \in \cS} [ f_j^2 T(f_j) ] \Big\rag_{\rC}
      \Big\lag \prod_{j \in \bar\cS} [ \bar f_j^2 \bar T(\bar f_j) ] \Big\rag_{\rC} \Big], ~~
f_j = \ep^{\f{\pi\ii j}{p}}, ~~ \bar f_j = \ep^{-\f{\pi\ii j}{p}},
\ee
where the sum $\cS$ is over all the subsets of $\cS_0=\{0,1,\cdots,2p-1\}$, including the empty set $\vn$ and $\cS_0$ itself, and $\bar \cS$ is the complement set $\bar\cS=\cS_0/\cS$.
First one needs to evaluate the right hand side of (\ref{TheCoefficient}) for a general positive integer $p$ and then take the analytic continuation $p \to \f12$.
Unfortunately, we do not know how to evaluate $y_T$.
In the following we will fit it numerically from the special case $D(\r_A(\vn),\r_A(L))$ in the spin chain results and check the coefficient in the other cases.
Since the OPE of twist operators has been used, in order the equation (\ref{TheFormula}) being valid we need that the interval length $\ell$ be much smaller than any characteristic length of the two states $\cL$, i.e. $\ell \ll \cL$, which includes both the size of the total system $L$ and the inverse temperature $\b$.

\begin{figure}[htbp]
  \centering
  \includegraphics[height=0.5\textwidth]{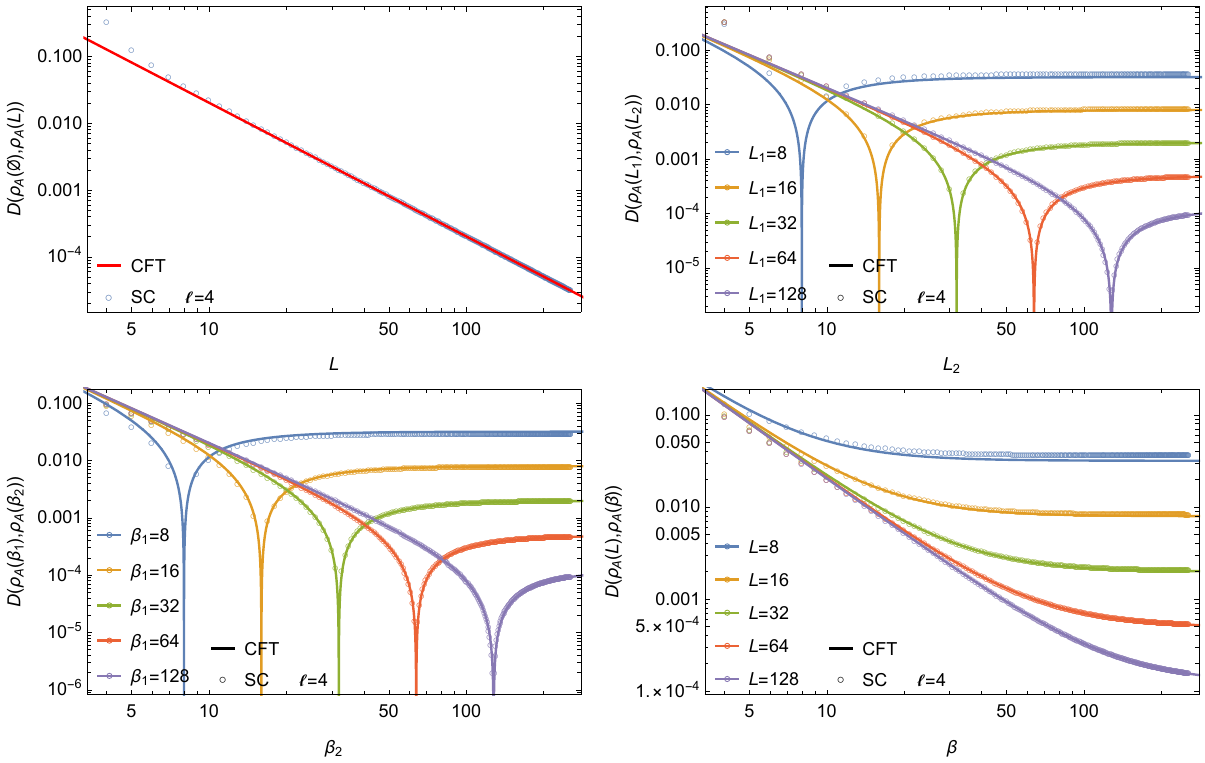}\\
  \caption{Trace distance of the RDMs in states on the cylinders in the free massless Majorana fermion theory (solid lines) and the critical Ising chain (empty circles).}\label{IsingTD}
\end{figure}

In the ground state on a circle $\r(L)$ we have that the expectation value of the stress tensor reads
\be
\lag T \rag_{\r(L)} =  \f{\pi^2 c}{6L^2}.
\ee
Combining both the CFT and spin chain results, we get
\be \label{IsingesvsL}
D(\r_A(\vn),\r_A(L)) \approx 0.126 \f{\ell^2}{L^2} + o\Big(\f{\ell^2}{L^2}\Big).
\ee
In CFT we know that the leading order trace distance is proportional to $\f{\ell^2}{L^2}$, and we obtain the approximate overall coefficient $0.126$ from numerical fit of the spin chain results.
This gives the approximate value of (\ref{TheCoefficient}) $y_T \approx 0.154$.%
\footnote{The formula (\ref{TheFormula}) also applies to the trace distance $D(\r_A(L),\r_{A,\ve}(L))$, with $\r_{A,\ve}(L)$ being the RDM of the energy eigenstate $\r_\ve(L)$. The state $\r_\ve(L)$ represents a vertical cylinder with spatial circumference $L$ and the operator $\ve$ being inserted at its two ends in the infinity.
In \cite{Zhang:2019itb} it was obtained numerically
\[ D(\r_A(L),\r_{A,\ve}(L)) \approx 0.153 \f{ 2\pi^2\ell^2}{L^2} + o\Big(\f{\ell^2}{L^2}\Big), \]
which gives $y_T \approx 0.153$.
Neither the value $y_T \approx 0.153$ in \cite{Zhang:2019itb} nor the value $y_T \approx 0.154$ is this paper is of high precision, mainly due to the small value of $L,\ell$.
In the following we will use $y_T \approx 0.154$ in the free massless Majorana fermion theory, which is precise enough for us in the paper.}
In the thermal state on an infinite line $\r(\b)$, we have the expectation values of the stress tensor
\be
\lag T \rag_{\r(\b)} =  - \f{\pi^2 c}{6\b^2}.
\ee
Based on (\ref{TheFormula}) and (\ref{IsingesvsL}), we further get
\be \label{IsingLvsL}
D(\r_A(L_1),\r_A(L_2)) \approx 0.126 \ell^2 \Big| \f{1}{L_1^2} - \f{1}{L_2^2} \Big| + o(\ell^2).
\ee
\be \label{Isingbetavsbeta}
D(\r_A(\b_1),\r_A(\b_2)) \approx 0.126 \ell^2 \Big| \f{1}{\b_1^2} - \f{1}{\b_2^2} \Big| + o(\ell^2).
\ee
\be \label{IsingLvsbeta}
D(\r_A(L),\r_A(\b)) \approx 0.126 \ell^2 \Big( \f{1}{L^2} + \f{1}{\b^2} \Big) + o(\ell^2).
\ee
Some of the results are plotted in Fig.~\ref{IsingTD}. We see perfect matches of the CFT and spin chain results for $\ell/\cL\ll1$ with $\cL$ being all values of $L$ and $\b$.

When at least one of the two states $\r,\s$ are on a torus with $\lag \ve \rag_\r \neq \lag \ve \rag_\s$, the leading order short interval expansion of the trace distance is \cite{Zhang:2019wqo,Zhang:2019itb}
\be \label{DrAsAMajorana1}
D(\r_A,\s_A) = \f{\ell}{2\pi} | \lag \ve \rag_\r - \lag \ve \rag_\s | + o(\ell).
\ee
However, when $|\lag \ve \rag_\r - \lag \ve \rag_\s|$ is exponentially small while $|\lag T \rag_\r - \lag T \rag_\s|$ is not, the dominate contribution to the trace distance would be (\ref{TheFormula}).
When the terms (\ref{DrAsAMajorana1}) and (\ref{TheFormula}) are at the same order, we do not have a reliable CFT result.
In the critical Ising chain, we could calculate numerically the trace distance for such states.
As we do no have reliable CFT results to be compared with, we will not show these spin chain results here.

\subsection{Schatten two-distance}

\begin{figure}[htbp]
  \centering
  \includegraphics[height=0.5\textwidth]{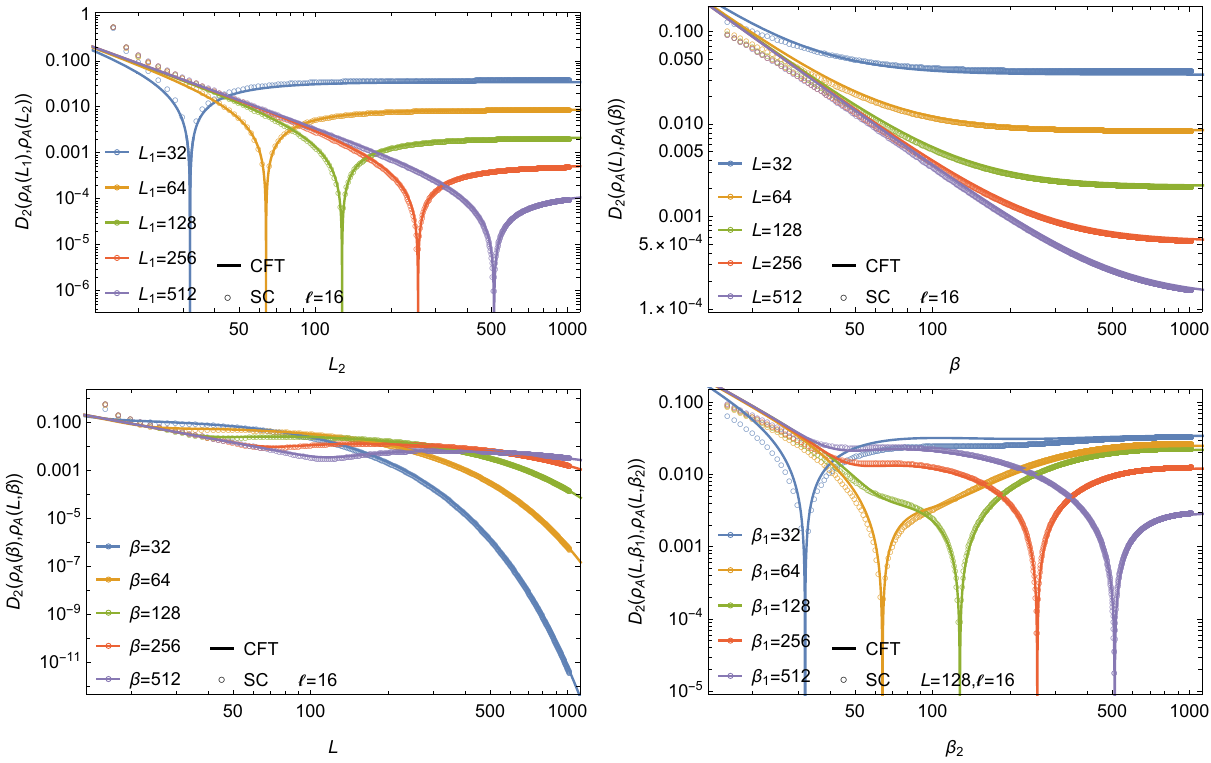}\\
  \caption{Schatten two-distance of the RDMs in states on the cylinders and toruses in the free massless Majorana fermion theory (solid lines) and the critical Ising chain (empty circles).}
  \label{IsingD2}
\end{figure}

We define the subsystem Schatten two-distance of two RDMs $\r_A,\s_A$ as
\be
D_2(\r_A,\s_A) = \sr{\f{\tr_A(\r_A-\s_A)^2}{2\tr_A(\r_A(\vn)^2)}}.
\ee
Note that in the ground state of the 2D CFT on the plane \cite{Holzhey:1994we,Calabrese:2004eu}
\be
\tr_A(\r_A(\vn)^2) = c_2 \Big(\f{\ell}{\e}\Big)^{-2\D_2},
\ee
with scaling dimension for the twist operators \cite{Calabrese:2004eu}
\be
\D_n = \f{c(n^2-1)}{12n}.
\ee
We have normalized the Schatten two-distance so that it is scale invariant and does not depend on the UV cutoff.
Short interval expansion of Schatten two-distance could be calculated from the OPE of twist operators \cite{Basu:2017kzo,He:2017txy}.
For the finite size and thermal states, including states on the plane, cylinders and toruses, we get
\be \label{D2Ising}
D_2(\r_A,\s_A) = \f{1}{16} \sr{ 8 \ell^2 ( \lag \ve \rag_\r - \lag \ve \rag_\s )^2 + 7 \ell^4 ( \lag T \rag_\r - \lag T \rag_\s )^2 + O(\ell^6)}.
\ee
Note that $\lag T \rag_\r=\lag \bar T \rag_\s$ and the contributions from both the homomorphic and the anti-holomorphic sectors have been included.
As in the case of the R\'enyi entropy, we do not need the explicit RDMs to calculate the Schatten distance in spin chains, and correlation matrices are enough.
This allows us to compute the Schatten two-distance for a relatively large $\ell$ and compare it with the CFT results in Fig.~\ref{IsingD2}.

\subsection{Relative entropy}

For two density matrices $\r,\s$ the relative entropy is defined as
\be \label{REdef}
S(\r\|\s) = \tr (\r \log \r) - \tr (\r \log \s).
\ee
The replica trick to calculate the subsystem relative entropy in a 2D CFT was developed in \cite{Lashkari:2014yva,Lashkari:2015dia}.
For RDMs on the cylinders, there are analytical CFT results \cite{Sarosi:2016atx} which are valid for an interval $A=[0,\ell]$ with an arbitrary length
\bea \label{RECFT}
&& S(\r_A(L_1)\|\r_A(L_2)) = \f{c}{3} \log\f{L_2\sin\f{\pi\ell}{L_2}}{L_1\sin\f{\pi\ell}{L_1}}
                            +\f{c}{6} \Big(1-\f{L_2^2}{L_1^2}\Big) \Big( 1 - \f{\pi\ell}{L_2}\cot\f{\pi\ell}{L_2} \Big), \nn\\
&& S(\r_A(\b_1)\|\r_A(\b_2)) = \f{c}{3} \log\f{\b_2\sinh\f{\pi\ell}{\b_2}}{\b_1\sinh\f{\pi\ell}{\b_1}}
                              +\f{c}{6} \Big(1-\f{\b_2^2}{\b_1^2}\Big) \Big( 1 - \f{\pi\ell}{\b_2}\coth\f{\pi\ell}{\b_2} \Big), \nn\\
&& S(\r_A(L)\|\r_A(\b)) = \f{c}{3} \log\f{\b\sinh\f{\pi\ell}{\b}}{L\sin\f{\pi\ell}{L}}
                         +\f{c}{6} \Big(1+\f{\b^2}{L^2}\Big) \Big( 1 - \f{\pi\ell}{\b}\coth\f{\pi\ell}{\b} \Big), \nn\\
&& S(\r_A(\b)\|\r_A(L)) = \f{c}{3} \log\f{L\sin\f{\pi\ell}{L}}{\b\sinh\f{\pi\ell}{\b}}
                         +\f{c}{6} \Big(1+\f{L^2}{\b^2}\Big) \Big( 1 - \f{\pi\ell}{L}\cot\f{\pi\ell}{L} \Big).
\eea

\begin{figure}[htbp]
  \centering
  \includegraphics[height=0.5\textwidth]{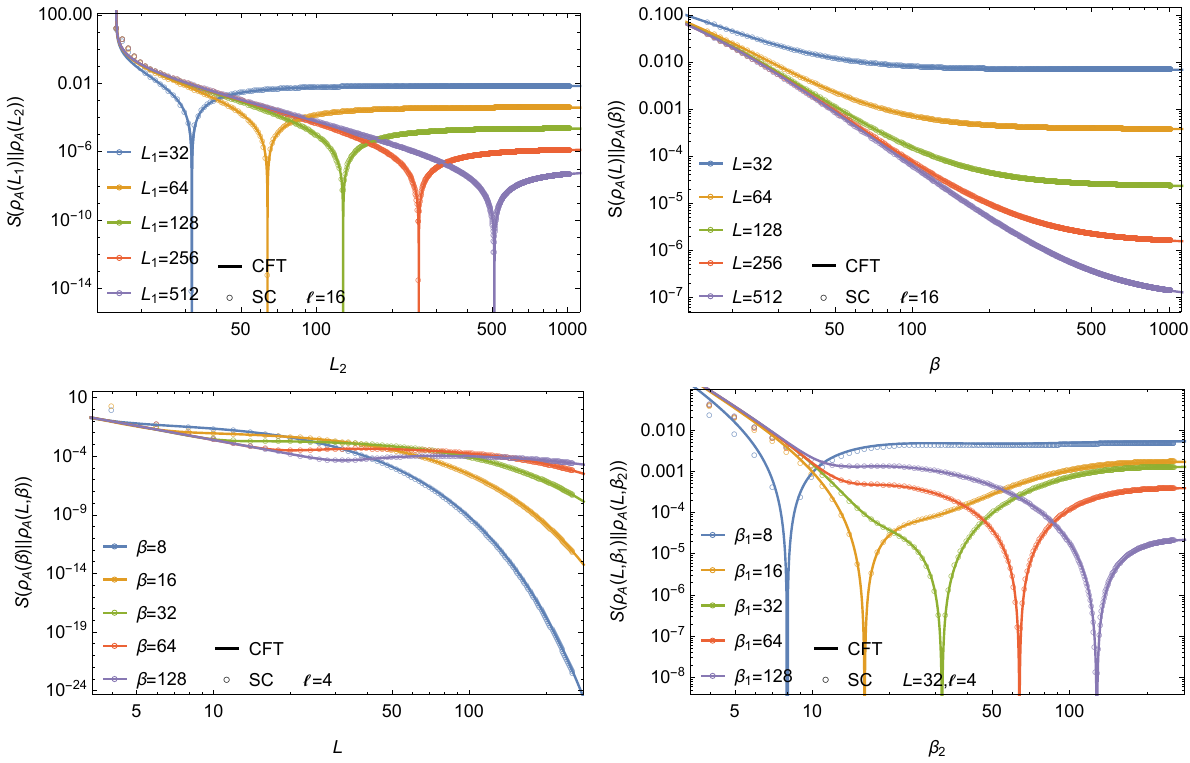}\\
  \caption{Relative entropy of the RDMs in states on the cylinders and toruses in the free massless fermion theory (solid lines) and the critical Ising chain (empty circles).}\label{IsingRE}
\end{figure}

For two Gaussian sates in the spin chain, the subsystem relative entropy \cite{Caputa:2016yzn} can be written in terms of the correlation matrix $\G$ defined in (\ref{Grho})
\be \label{PR}
S(\r_{\G_1}\|\r_{\G_2}) = \tr \Big( \f{1+\G_1}{2} \log \f{1+\G_1}{2} \Big) - \tr \Big( \f{1+\G_1}{2} \log \f{1+\G_2}{2} \Big),
\ee
This means that we just need to compute the $2\ell\times2\ell$ correlation matrix $\G$, rather than the explicit $2^\ell\times2^\ell$ RDM $\r_\G$, to obtain the relative entropy which allows us to check the CFT analytical results (\ref{RECFT}) for a long interval.
We show some of them in the top panels of Fig.~\ref{IsingRE}.
As the CFT results are exact, there are matches of the CFT and spin chain results not only for short intervals with $\ell \ll L$, $\ell \ll \b$ but also for long intervals with $\ell \sim L$, $\ell \sim \b$.

For the RDMs on the toruses we have to take the short interval expansion of the relative entropy from the OPE of twist operators.%
\footnote{Subsystem relative entropy on a torus could also be calculated from modular Hamiltonian \cite{Fries:2019acy}, which we will consider in this paper.}
The method of was developed in \cite{He:2017txy} following the replica trick in \cite{Lashkari:2014yva,Lashkari:2015dia}, and we get the following result for the RDMs on the toruses
\bea
&& S(\r_A\|\s_A) = \f{\ell^2}{12} ( \lag \ve \rag_\r - \lag \ve \rag_\s  )^2
               +\f{2\ell^4}{15} ( \lag T \rag_\r - \lag T \rag_\s  )^2 \nn\\
&& \phantom{S(\r_A\|\s_A) =}
               +\f{\ell^4}{15} ( \lag \ve \rag_\r - \lag \ve \rag_\s  )
                \big[ \lag T \rag_\r ( \lag \ve \rag_\r + \lag \ve \rag_\s ) - 2 \lag T \rag_\s \lag \ve \rag_\s \big] \nn\\
&& \phantom{S(\r_A\|\s_A) =}
               +\f{\ell^4}{120} ( \lag \ve \rag_\r - \lag \ve \rag_\s  )^2
                                \big( \lag \ve \rag_\r^2 + 2 \lag \ve \rag_\r \lag \ve \rag_\s + 3 \lag \ve \rag_\s^2 \big)
               +O(\ell^6).
\eea
In the critical Ising chain the state with both a finite size and a finite temperature is not Gaussian, and we cannot use the formula (\ref{PR}) to calculate the relative entropy in the spin chain.
In that case we need to construct explicitly the numerical RDMs and calculate the relative entropy from the definition (\ref{REdef}).
We compare the CFT and spin chain results in bottom panels of Fig.~\ref{IsingRE}.

\section{XX chain with zero transverse field}\label{secXX}

In this section we consider the XX chain with zero transverse field, and as was mentioned before its continuum limit gives the 2D free massless Dirac fermion theory, or equivalently the 2D free massless compact boson theory with unit target space radius, which is a 2D CFT with central charge $c=1$.
The calculations and results are parallel to those in the critical Ising chain and the 2D free massless Majorana fermion theory, and we will keep it brief in this section.

\subsection{von Neumann and R\'enyi entropies}





Details of the 2D free massless Dirac fermion theory and the 2D free massless compact boson theory could be found in \cite{DiFrancesco:1997nk,Blumenhagen:2009zz}.
In the NS sector of the 2D free massless Dirac fermion theory there are nonidentity primary operators
\be
J = \ii \psi_1 \psi_2, ~~ \bar J = \ii \bar \psi_1 \bar \psi_2, ~~ K = J \bar J,
\ee
with conformal weights $(1,0)$, $(0,1)$ and $(1,1)$ respectively.
In the R sector there are primary operators $\s_1$, $\s_2$ with the same conformal weights $(\f18,\f18)$.
In the NS and R sectors, there are also other primary operators with larger conformal weights, which are irrelevant to our low order computations in this paper.


The exact R\'enyi entropy for the interval $A=[0,\ell]$ on a torus with spatial circumference $L$ and temporal period $\b$ is \cite{Mukhi:2017rex}
\be \label{REmmwDirac}
S_A^{(n)} = \f{n+1}{6n}\log\Big| \f{L}{\e} \f{\th_1(x|\t)}{\th_1'(0|\t)} \Big|
           -\f{1}{n-1} \log\bigg[ \f{\sum_{\va,\vb}\Big| \Thab(0|\O) \Big|^2}
                                    {\big(\prod_{k=1}^{n-1}|A_k|\big)\big( \sum_{\n=2}^4|\th_\n(0|\t)|^2 \big)^n } \bigg].
\ee
Again we have defined $x=\f{\ell}{L}$, $\t=\ii\f{\b}{L}$ and the rest of the functions involved are in (\ref{Odef}), (\ref{ABodef}).
One could also see the R\'enyi entropy of one interval on a torus in the 2D free massless compact boson theory in \cite{Chen:2015cna}.

\begin{figure}
  \centering
  \includegraphics[height=0.25\textwidth]{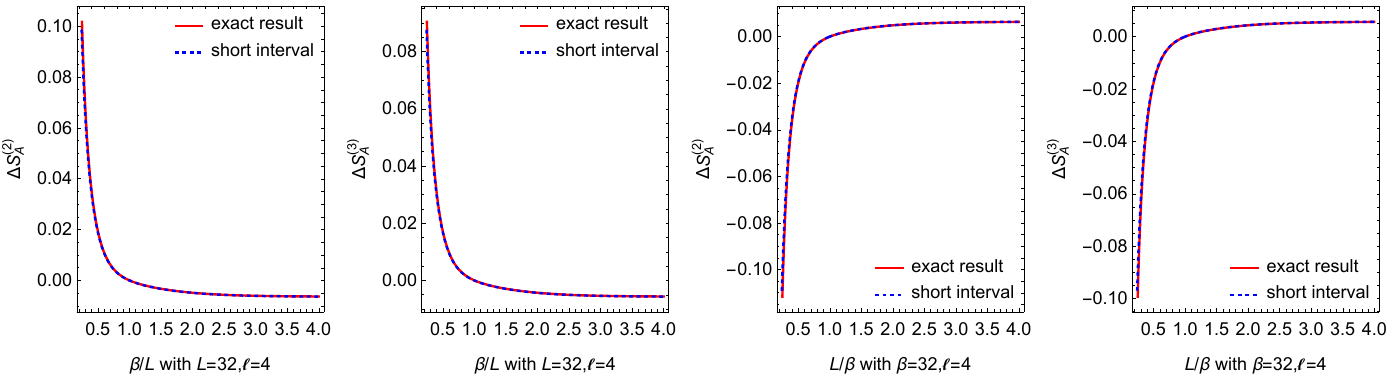}\\
  \caption{The comparison of the exact R\'enyi entropy with the short interval expansion one in the free massless Dirac fermion theory.
  We use $\D S_A^{(n)} = S_A^{(n)} - \f{n+1}{6n}\log\f{\ell}{\e}$ to eliminate the dependence on the UV cutoff.}
  \label{DiracRE2RE3}
\end{figure}

From OPE of twist operators we get the short interval expansion of the R\'enyi entropy on a torus
\be \label{REsieDirac}
S_A^{(n)}=\f{n+1}{6n}\log\f{\ell}{\e} - \f{(n+1)\ell^2}{6n}\lag T \rag + O(\ell^4),
\ee
with the expectation value
\be \label{TevDirac}
\lag T \rag = - \f{2\pi^2 q}{L^2} \p_q \log Z(q) , ~~
Z(q) = \f{1}{2\eta(\t)^2}\big[ \th_2(0|\t)^2+\th_3(0|\t)^2+\th_4(0|\t)^2 \big].
\ee
Note $q=\bar q=\ep^{-2\pi\b/L}$. The contributions from $\bar T$ have also been included.
Short interval expansion of the exact result (\ref{REmmwDirac}) gives
\be
S_A^{(n)} = \f{n+1}{6n}\log\f{\ell}{\e}
          + \f{(n+1)\ell^2}{12nL^2}
           \Big( \f13 \f{\th_1'''(0|\t)}{\th_1'(0|\t)}
                -\f{\sum_{\n=2}^4\th_\n(0|\t)\th_\n''(0|\t)}{\sum_{\n=2}^4\th_\n(0|\t)^2}
           \Big) + O(\ell^{4}),
\ee
which is the same as the short interval expansion result from twist operators (\ref{REsieDirac}).
This is an indication that the method of short interval expansion from the OPE of twist operators is valid to order $\ell^2$, but as we show in appendix~\ref{appBDT} the method fails to give the correct R\'enyi entropy at order $\ell^4$.
We compare the exact R\'enyi entropy and short interval result in Fig.~\ref{DiracRE2RE3}.
We see that the short interval expansion for the R\'enyi entropy is a good approximation in the parameter regimes we consider.
Taking $n \to 1$ limit for the R\'enyi entropy (\ref{REsieDirac}), we get the short interval expansion of the von Neumann entropy
\be \label{EEsieDirac}
S_A = \f{1}{3}\log\f{\ell}{\e} - \f{\ell^2}{3} \lag T \rag
+ O(\ell^6).
\ee

\begin{figure}[htbp]
  \centering
  \includegraphics[height=0.4\textwidth]{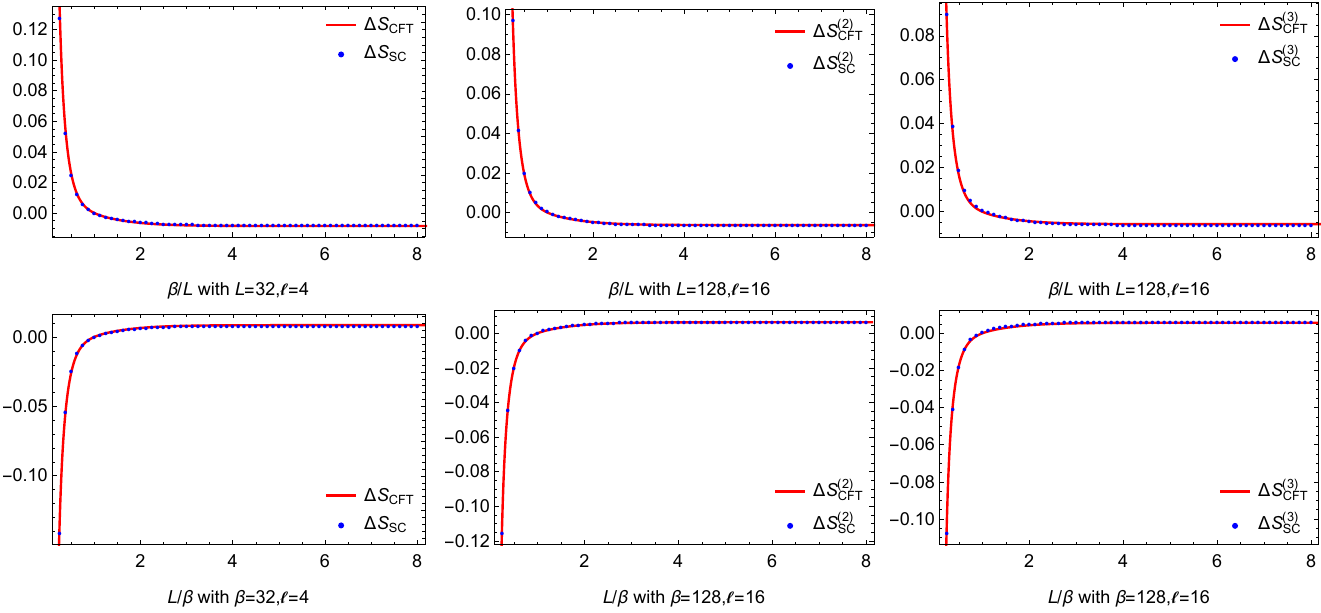}\\
  \caption{We compare the von Neumann and R\'enyi entropies in the free massless Dirac fermion theory and those in the XX chain with zero field.}
  \label{XXEERE2RE3}
\end{figure}

In the XX chain with zero field, we construct numerically the RDMs of one interval in the finite size and thermal states as detailed in appendix~\ref{appRDM}.
We study the XX chain with a total number of sites $L$, that is four times of an integer. As there are two zero modes in the R sectors we will need to use again the trick developed in \cite{Fagotti:2010yr}.
We compute the von Neumann entropy for a short interval from the explicit numerical RDM, and calculate the R\'enyi entropy for a relatively long interval from the correlation matrices.
We compare the CFT and spin chain results in Fig.~\ref{XXEERE2RE3}.
On the CFT side, we use the short interval expansion of the von Neumann entropy (\ref{EEsieDirac}) and the exact R\'enyi entropy (\ref{REmmwDirac}).
We see perfect matching between the CFT and spin chain results.

\subsection{Trace distance}

\begin{figure}[htbp]
  \centering
  \includegraphics[height=0.5\textwidth]{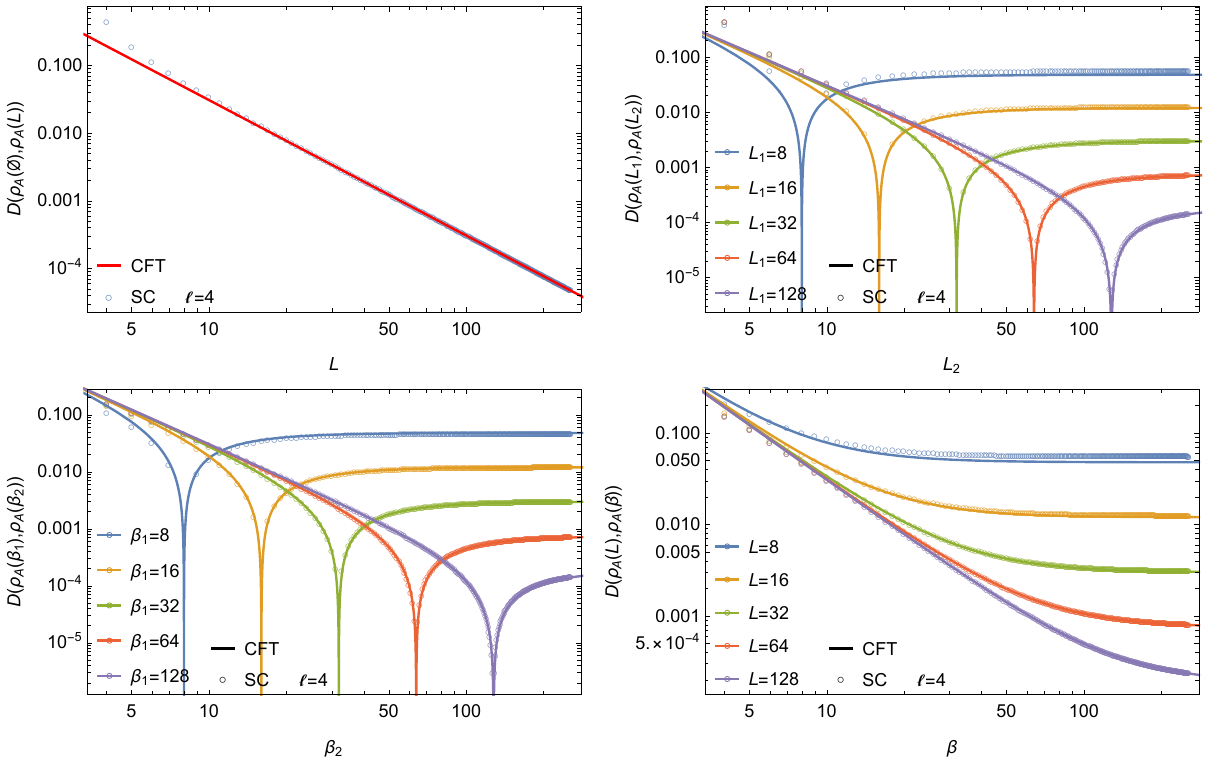}\\
  \caption{Trace distance of the RDMs on the cylinders in the free massless Dirac fermion theory (solid lines) and the XX chain with zero field (empty circles).}\label{XXTD}
\end{figure}

We compute the trace distance among the RDMs in states on the plane and cylinders in the 2D free massless Dirac fermion theory.
The trace distance $D(\r_A(\vn),\r_A(L))$ can be written as (\ref{TheFormula}) with the coefficient (\ref{TheCoefficient}) that we cannot evaluate in the CFT.
By fitting of the numerical results in the XX chain with $\ell=4$, we obtain the trace distance
\be \label{XXesvsL}
D(\r_A(\vn),\r_A(L)) \approx 0.191 \f{\ell^2}{L^2} + o\Big(\f{\ell^2}{L^2}\Big),
\ee
which gives the approximate coefficient $y_T \approx 0.164$.%
\footnote{In the 2D free massless Dirac fermion theory, the formula (\ref{TheFormula}) also applies to the trace distance $D(\r_A(L),\r_{A,K}(L))$, with $\r_{A,K}(L)$ being the RDM of the energy eigenstate $\r_K(L)$.
In \cite{Zhang:2019itb} it was obtained numerically
\[ D(\r_A(L),\r_{A,K}(L)) \approx 0.166 \f{ 2\sr{2}\pi^2\ell^2}{L^2} + o\Big(\f{\ell^2}{L^2}\Big), \]
which gives $y_T \approx 0.166$. Neither the value $y_T \approx 0.166$ in \cite{Zhang:2019itb} nor the value $y_T \approx 0.164$ is this paper is of high precision, due to the small values of $L,\ell$.}
We will use this approximate value in the free massless Dirac fermion theory.
For the RDMs of one interval in states on the cylinders we get
\bea
&& D(\r_A(L_1),\r_A(L_2)) \approx 0.191 \ell^2 \Big| \f{1}{L_1^2} - \f{1}{L_2^2} \Big| + o(\ell^2), \nn\\
&& D(\r_A(\b_1),\r_A(\b_2)) \approx 0.191 \ell^2 \Big| \f{1}{\b_1^2} - \f{1}{\b_2^2} \Big| + o(\ell^2), \nn\\
&& D(\r_A(L),\r_A(\b)) \approx 0.191 \ell^2 \Big( \f{1}{L^2} + \f{1}{\b^2} \Big) + o(\ell^2).
\eea
These analytical CFT results and numerical spin chain results are compared in Fig.~\ref{XXTD}.

For two states $\r,\s$ on the torus, there are generally three quasiprimary operators at level two $K$, $T$, $\bar T$ that have different expectation values. Using the method in \cite{Zhang:2019wqo,Zhang:2019itb}, we cannot calculate the trace distance among the RDMs on the torus in the free massless Dirac fermion theory.
As there are no CFT results to be compared with, we will not show the trace distance involving the RDMs in states with both finite sizes and finite temperatures in the XX chain in this paper.

\subsection{Schatten two-distance}

\begin{figure}[htbp]
  \centering
  \includegraphics[height=0.5\textwidth]{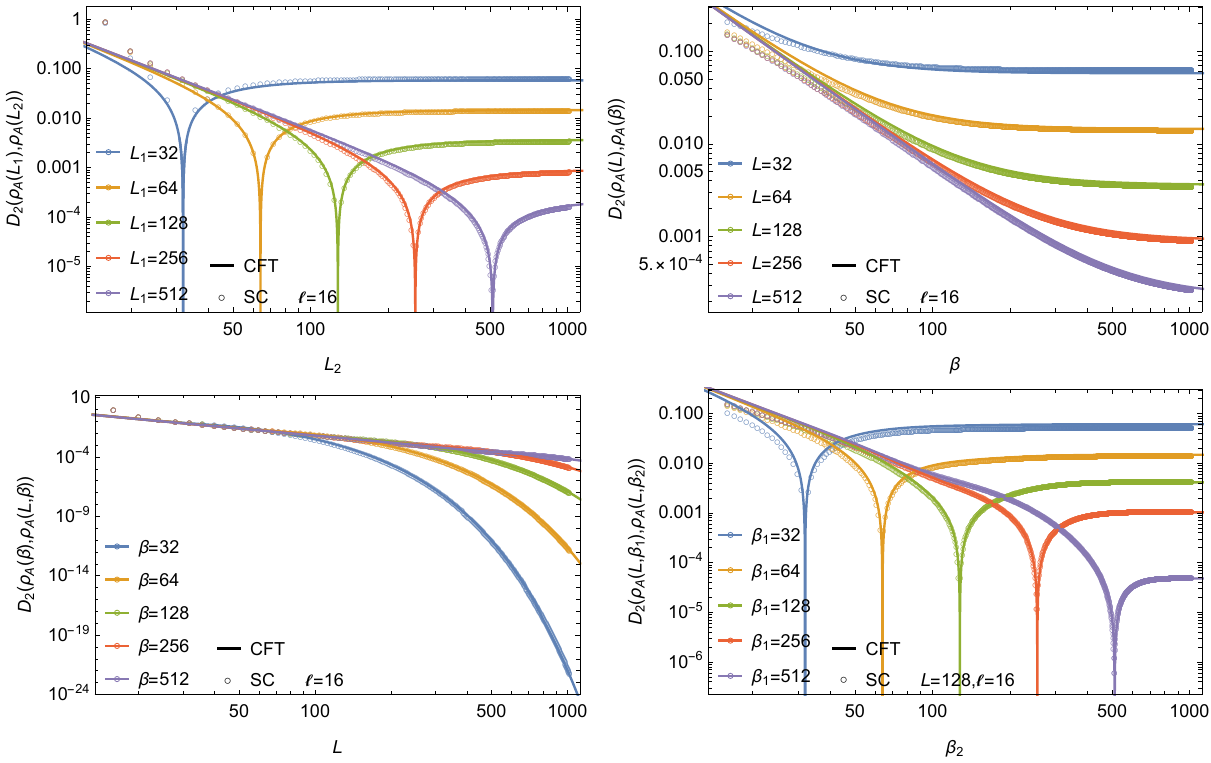}\\
  \caption{Schatten two-distance of the RDMs on the cylinders and toruses in the free massless Dirac fermion theory (solid lines) and the XX chain with zero field (empty circles).}
  \label{XXD2}
\end{figure}

In the free massless Dirac fermion theory we get the short interval expansion of the Schatten two-distance from the OPE of twist operators
\be
D_2(\r_A,\s_A) = \f{\ell^2}{16\sr{2}} \sr{ ( \lag K \rag_\r - \lag K \rag_\s )^2 +  10 ( \lag T \rag_\r - \lag T \rag_\s )^2 + O(\ell^2)}.
\ee
Note that on a torus with $q=\bar q=\ep^{2\pi\ii\t}=\ep^{-2\pi\b/L}$ we have the expectation value of stress tensor (\ref{TevDirac}) and
\be \label{KevDirac}
\lag K \rag = \f{4\pi^2}{L^2} q \p_q \log \f{\th_3(0|2\t)}{\th_3(0|\t/2)}.
\ee
The contributions from $\bar T$ have also been included.
We compare the analytical results of the Schatten two-distance in the free massless Dirac fermion theory and the numerical results in the XX chain with zero field in Fig.~\ref{XXD2}.

\subsection{Relative entropy}

\begin{figure}[htbp]
  \centering
  \includegraphics[height=0.5\textwidth]{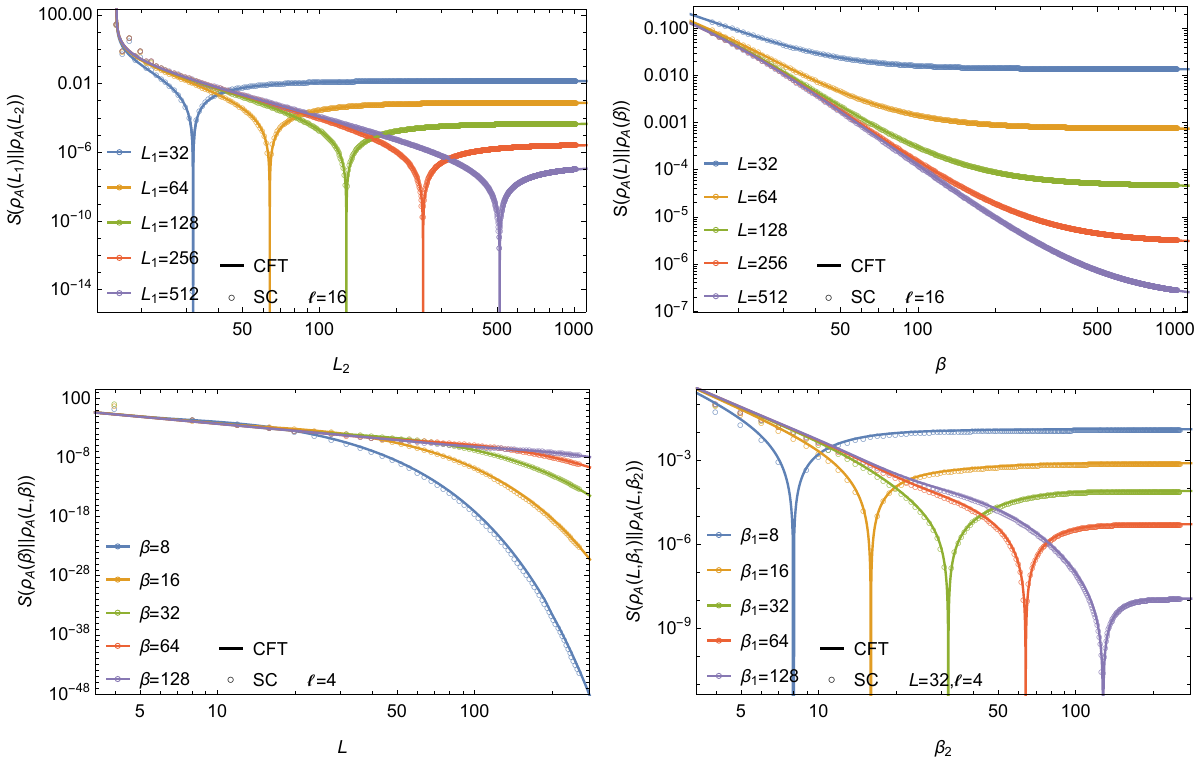}\\
  \caption{Relative entropy of the RDMs on the cylinders and toruses in the free massless Dirac fermion theory (solid lines) and XX chain with zero field (empty circles).}
  \label{XXRE}
\end{figure}

The results of relative entropy of RDMs on the cylinders (\ref{RECFT}) are universal and apply to any 2D CFT.
For RDMs on the toruses, we get the short interval expansion of the relative entropy from the OPE of twist operators
\be
S(\r_A\|\s_A) = \f{\ell^4}{60} ( \lag K \rag_\r - \lag K \rag_\s )^2 + \f{\ell^4}{15} ( \lag T \rag_\r - \lag T \rag_\s )^2 + O(\ell^6),
\ee
with the expectation values (\ref{TevDirac}), (\ref{KevDirac}).
The contributions from the anti-holomorphic sector have been included.
We compare the CFT and spin chain results in Fig.~\ref{XXRE}.

\section{Conclusion and discussion}\label{secCon}

In this paper, we have constructed the numerical RDMs of an interval in the finite size and thermal states in the critical XY chains, specially for the states with both a finite size and a finite temperature, focusing on the critical Ising chain and the XX chain with zero transverse field.
With the numerical RDMs, we computed the subsystem von Neumann entropy, R\'enyi entropy, trace distance, Schatten two-distance, and relative entropy, and compared the results with those in the 2D free massless Majorana and Dirac fermion theories, which are respectively the continuum limits of the critical Ising chain and the XX chain with zero field.
We found perfect matches of the numerical spin chain and analytical CFT results in the continuum limit.

There are several interesting generalizations of the present results.
In CFT, we only got short interval expansion of von Neumann entropy of a length $\ell$ interval to order $\ell^2$, and it is interesting to calculate higher order results.
We cannot calculate subsystem trace distance for RDMs in states with both a finite size and a finite temperature in CFT, and other methods to calculate the subsystem trace distance are needed.
The states with both a finite size and a finite temperature in the XY spin chains are not Gaussian, and we can only calculate the von Neumann entropy, trace distance and relative entropy for a short interval. It would be interesting to calculate those quantities for a long interval in spin chains.

We have only calculated the results numerically in the spin chain, and it is interesting to calculate the spin chain results analytically, like for example the ground state entanglement entropy and R\'enyi entropy in \cite{jin2004quantum}.
The analytical calculations would be difficult if possible.
For some quantities, like the R\'enyi entropy and the Schatten distance, we need to manipulate the $2\ell\times2\ell$ correlation matrices, some of which are not of the Toeplitz type, and this makes the analytical calculations difficult.
For other quantities, like the von Neumann entropy and the trace distance, we need to manipulate the $2^\ell\times2^\ell$ RDMs, and they are more difficult to calculate analytically than the R\'enyi entropy and Schatten distance.

We have elaborated on how to calculate the subsystem distances among the finite size and thermal states in CFTs and spin chains. As we stated above, some of the results are very limited.
It would be interesting to develop new techniques and obtain more general results, for which there are many potential applications.
One potential application of these results is to investigate the thermalization of subsystems in a finite total system, like that in \cite{fagotti2013reduced} for thermalization of subsystems in an infinite total system after a global quantum quench \cite{Calabrese:2005in,Calabrese:2006rx,Calabrese:2016xau}. Another possible application is the distinguishability of the black hole microstates and other states in gravity and holographic CFTs \cite{Bao:2017guc,Michel:2018yta,Guo:2018djz,Dong:2018lsk,Guo:2018fye}.

\section*{Acknowledgements}

We thank Pasquale Calabrese and Erik Tonni for helpful discussions, comments, and suggestions.
JZ acknowledges support from ERC under Consolidator grant number 771536 (NEMO).

\appendix

\section{Break down of the method of twist operators at order $\ell^4$}\label{appBDT}

In this appendix, we show that the method of OPE of twist operators cannot give the correct short interval R\'enyi entropy on a torus at order $\ell^4$ in some specific 2D CFTs, including the 2D free massless Majorana and Dirac fermion theories.

In a general 2D unitary CFT, we consider the nonidentity primary operators $\phi_i$, $i=1,2,\cdots,g$ with the smallest scaling dimension $\D$.
There is a degeneracy $g$ at scaling dimension $\D$ and each primary operator $\phi_i$ has the conformal weights $(h_i,\bar h_i)$. Note that $\D=h_i+\bar h_i$ for all $i$.
We require that $0<\D<2$ and at least one of these primary operator $\phi_i$ is non-chiral, i.e. both $h_i>0$ and $\bar h_i>0$.
Apparently, the 2D free massless Majorana and Dirac fermion theories belong to such theories.
For the 2D free massless Majorana fermion theory, the operator is $\s$ with conformal weights $(\f{1}{16},\f{1}{16})$, and there is no degeneracy $\D=\f{1}{8}$, $g=1$.
For the 2D free massless Dirac fermion theory, the operators are $\s_1,\s_2$ with the same conformal weights $(\f{1}{8},\f{1}{8})$, and there is double degeneracy $\D=\f{1}{4}$, $g=2$.

We consider the R\'enyi entropy of one interval $A=[0,\ell]$ in the 2D CFT on a torus with spatial circumference $L$ and temporal period $\b$.
In the low temperature limit $L\ll\b$, the density matrix of the whole system could be written as an expansion in the variable $q=\ep^{-2\pi\b/L}$
\be \label{staterho}
\r = \f{|0\rag \lag 0 | + q^\D \sum_{i=1}^g |\phi_i\rag \lag \phi_i | +o(q^\D) }{1+g q^\D+o(q^\D)}.
\ee
We have the ground state $|0\rag$ and the orthonormal primary excited states $| \phi_i \rag$ that satisfy $\lag \phi_i | \phi_j \rag=\d_{ij}$.
There is a universal single interval R\'enyi entropy  \cite{Cardy:2014jwa} in this case that reads
\be \label{RenyiLTX}
S_A^{(n)} = \f{c(n+1)}{6n} \log \Big( \f{L}{\pi\e} \sin\f{\pi\ell}{L} \Big)
           -\f{n g q^\D}{n-1} \Big[ \f{1}{n^{2\D}}\Big(\f{\sin\f{\pi\ell}{L}}{\sin\f{\pi\ell}{nL}}\Big)^{2\D} -1\Big]
           +o(q^\D).
\ee

To compare, we can compute the same R\'enyi entropy using OPE of twist operators. In general one has \cite{Chen:2016lbu}
\bea \label{RenyiSIX}
&& S_A^{(n)} = \f{c(n+1)}{6n} \log \f{\ell}{\e}
           -\f{1}{n-1}
           \log
            \Big\{
              1
             + \ell^2 b_T ( \lag T \rag + \lag \bar T \rag ) \nn\\
&& \phantom{S_A^{(n)} =}
             + \ell^4 [ b_\cA ( \lag \cA \rag + \lag \bar \cA \rag ) + b_{TT}( \lag T \rag^2 + \lag \bar T \rag^2 ) + b_T^2 \lag T \rag\lag \bar T \rag  ]
             +O(\ell^6) \nn\\
&& \phantom{S_A^{(n)} =}
             + \sum_{\psi} [ \ell^{2\D_\psi} b_{\psi\psi} \lag \psi \rag^2 + O(\ell^{2\D_\psi}) ]
             \Big\},
\eea
with the coefficients
\be
b_T=\frac{n^2-1}{12n}, ~~
b_\mA=\frac{(n^2-1)^2}{288 n^3}, ~~
b_{TT}=\frac{(n^2-1) [5 c (n+1) (n-1)^2+2 (n^2+11)]}{1440 c n^3},
\ee
and the level four quasiprimary operator
\be
\cA = (TT) - \f{3}{10} \p^2 T.
\ee
It is similar for the anti-holomorphic quasiprimary operators $\bar T$, $\bar \cA$.
The sum $\psi$ is over all the nonidentity primary operators in the theory.
The following argument show the coefficient $b_{\psi\psi}$ will be irrelevant at the order of the expansion we are interested in.
In state (\ref{staterho}), we have that the expectation value for an arbitrary operator $\cO$
\be \label{evO}
\lag \cO \rag =  \lag \cO \rag_0 +  q^\D \sum_{i=1}^g ( \lag \cO \rag_{\phi_i} - \lag \cO \rag_0 )  + o(q^\D).
\ee
with $\lag \cO \rag_0$ being the expectation value in the ground state $|0\rag$ and $\lag \cO \rag_{\phi_i}$ being the one in the primary excited state $|\phi_i\rag$.
On the torus in the low temperature limit, for a primary operator $\psi$ there is a leading order expectation value $\lag \psi \rag \sim q^\D$.
As we will focus on the order $q^\D$ part of the R\'enyi entropy, we do not need to consider the contributions from the nonidentity primary operators, i.e. the terms with $\psi$ in (\ref{RenyiSIX}).

On a torus in the low temperature limit $q\ll1$, using (\ref{evO}) and $\lag T \rag_{\phi_i},\lag \cA \rag_{\phi_i}$ in \cite{Lin:2016dxa} we get the expectation values
\bea
&& \hspace{-10mm}
   \lag T \rag = \frac{\pi ^2 [c-24 H q^{\Delta } + o(q^\D)  ]}{6 L^2}, ~~
   \lag \cA \rag = \frac{\pi ^4 [ c (5 c+22)-240 q^{\Delta } ((c+2) H - 12 H_2) + o(q^\D) ]}{180 L^4}, \nn\\
&& \hspace{-10mm}
   \lag \bar T \rag = \frac{\pi ^2 [c-24 \bar H q^{\Delta } + o(q^\D)  ]}{6 L^2}, ~~
   \lag \bar \cA \rag = \frac{\pi ^4 [ c (5 c+22)-240 q^{\Delta } ((c+2) \bar H - 12 \bar H_2) + o(q^\D) ]}{180 L^4},
\eea
with the following definitions
\be
H=\sum_{i=1}^g h_i, ~~
H_2=\sum_{i=1}^g h_i^2, ~~
\bar H=\sum_{i=1}^g \bar h_i, ~~
\bar H_2=\sum_{i=1}^g \bar h_i^2.
\ee

We compare the low temperature expansion of the R\'enyi entropy (\ref{RenyiLTX}) with the short interval expansion result (\ref{RenyiSIX}) and
focus on the order $q^\D$ part of the R\'enyi entropy.
At order $\ell^2$, they are the same but at order $\ell^4$, there is the non-vanishing difference
\be
\frac{\pi ^4 (n-1) (n+1)^2 ( H_2+\bar H_2 - g \Delta^2 )\ell ^4 q^{\Delta }}{18 n^3 L^4}.
\ee
It is essential that the lightest nonidentity primary operators have a scaling dimension $0<\D<2$ and at least one of lightest nonidentity primary operator is non-chiral.
This is consistent with the result in \cite{Mukhi:2017rex,Mukhi:2018qub} where the authors argued that the twist operators cannot give the correct R\'enyi entropy in the 2D free massless fermion theories on a torus.
We have shown that the method of OPE of twist operators breaks down in more general 2D CFTs on a torus.
In these 2D CFTs, the method of OPE of twist operators cannot give the correct R\'enyi entropy on a torus, but it is still possible that it could give the correct von Neumann entropy.
It is interesting to study whether it is the case or not.

We have only shown in what kind of 2D CFTs the method of short interval expansion from the OPE of twist operators for the torus R\'enyi entropy is not valid, but we do not have a criterion in what kind of theories the method is valid.
We have a sufficient but not necessary condition for the method being invalid.
For the theories that the condition is not satisfied, there may be conditions at higher orders that make the method invalid.
We only have constraints for the spectrum, but we have no constraint for the central charge. This is an interesting direction to be explored in the future.

Another related question is whether the method is valid in the holographic 2D CFT with a large central charge and a sparse spectrum \cite{Brown:1986nw,Strominger:1997eq,Hartman:2014oaa}.
In fact the method of twist operators has been used in \cite{Chen:2016lbu,Zhang:2016rja,He:2017txy} to calculate the torus R\'enyi entropy in the 2D large central charge CFT, using the spectrum of only the vacuum conformal family operators and other chiral operators. The results are the same as those computed from other methods in \cite{Barrella:2013wja,Cardy:2014jwa,Chen:2014unl,Chen:2015uia}.
If light nonchiral primary operators with scaling dimension $0<\D<2$ are included in the spectrum, one would meet the same problem as above.

\section{Thermal RDM in XY chains}\label{appRDM}

The spin-$\f12$ XY chain with transverse field has the Hamiltonian
\be \label{HS}
H_{\rm XY} = - \sum_{j=1}^L \Big( \f{1+\g}{4}\s_j^x\s_{j+1}^x + \f{1-\g}{4}\s_j^y\s_{j+1}^y + \f{\l}{2}\s_j^z \Big),
\ee
where $L$ is the total number of sites.
In this paper, we only consider the cases on which $L$ is multiple of four.
We consider the periodic boundary conditions $\s_{L+1}^{x,y,z}=\s_{1}^{x,y,z}$ for the Pauli matrices $\s_{j}^{x,y,z}$.
When $\g=\l=1$ it defines the critical Ising chain, and its continuum limit gives the 2D free massless Majorana fermion theory.
When $\g=\l=0$ it defines the XX chain with zero transverse field, and its continuum limit gives the 2D free massless Dirac theory, or equivalently the 2D free massless compact boson theory with the target space being a unit radius circle.
The Hamiltonian of the XY chain can be exactly diagonalized \cite{Lieb:1961fr,katsura1962statistical,pfeuty1970one} and the numerical RDMs in the ground state and excited energy eigenstates could be constructed following
\cite{chung2001density,Vidal:2002rm,peschel2003calculation,Latorre:2003kg,jin2004quantum,Alba:2009th,Alcaraz:2011tn,Berganza:2011mh}.
The construction of the RDM in a thermal state on an infinite line could be found in \cite{fagotti2013reduced}.
In this appendix, we elaborate on how to construct the numerical RDM of one interval in a state with both a finite size and a finite temperature.
Along the construction, the trick in \cite{Fagotti:2010yr} will be extremely useful to us.

The XY chain Hamiltonian can be exactly diagonalized by successively applying the Jordan-Wigner transformation, Fourier transforming, and Bogoliubov transformation.
The Jordan-Wigner transformation is
\be
a_j = \Big(\prod_{i=1}^{j-1}\s_i^z\Big) \s_j^+, ~~
a_j^\dag = \Big(\prod_{i=1}^{j-1}\s_i^z\Big) \s_j^-,
\ee
with $\s_j^\pm = \f12 ( \s_j^x \pm \ii \s_j^y )$.
In the NS sector there are antiperiodic boundary conditions $a_{L+1}=-a_1$, $a_{L+1}^\dag=-a_1^\dag$, and in the R sector there are periodic boundary conditions $a_{L+1}=a_1$, $a_{L+1}^\dag=a_1^\dag$.
The Fourier transformation is
\be
b_k = \f{1}{\sr{L}}\sum_{j=1}^L\ep^{\ii j \vph_k}a_j, ~~
b_k^\dag = \f{1}{\sr{L}}\sum_{j=1}^L\ep^{-\ii j \vph_k}a_j^\dag,
\ee
with $\vph_k = \f{2\pi k}{L}$.
The momenta $k$'s are half integers in the NS sector
\be
k =\f{1-L}{2}, \cdots, -\f12, \f12, \cdots, \f{L-1}{2},
\ee
and integers  in the R sector
\be
k =1-\f{L}{2}, \cdots, -1, 0, 1, \cdots, \f{L}{2}.
\ee
The Bogoliubov transformation is
\be
c_k = b_k \cos\f{\th_k}{2} + \ii b_{-k}^\dag \sin\f{\th_k}{2}, ~~
c_k^\dag = b_k^\dag \cos\f{\th_k}{2} - \ii b_{-k} \sin\f{\th_k}{2}.
\ee
For the critical Ising chain, we choose the angle
\be
\th_k =
\lt\{\ba{ll}
 -\f\pi2-\f{\pi k}{L} & k<0 \\
 0 & k=0 \\
 \f\pi2-\f{\pi k}{L}  & k>0
\ea\rt. \!\!\! .
\ee
For the XX chain, the Bogoliubov transformation is not needed, and, in other words, there is always $\th_k=0$.

Finally, the Hamiltonian becomes
\be
H =  \f{1+\cP}{2} H_{\rm NS}
   + \f{1-\cP}{2} H_{\rm R}, ~~
H_{\rm NS} = \sum_{k\in{\rm NS}} \ve_k \Big( c_k^\dag c_k -\f12 \Big), ~~
H_{\rm R} = \sum_{k\in{\rm R}} \ve_k \Big( c_k^\dag c_k -\f12 \Big).
\ee
In the critical Ising chain we have
\be
\ve_k = 2 \sin \f{\pi|k|}{L},
\ee
and in the XX chain with zero transverse field
\be
\ve_k = - \cos \f{2\pi k}{L}.
\ee
The projection operator is
\be
\cP = \prod_{j=1}^{L} \s_j^z = \ep^{\pi\ii\sum_{j=1}^L a_j^\dag a_j}.
\ee

One can define the Majorana modes as
\be
d_{2j-1} = a_j + a_j^\dag, ~~
d_{2j} = \ii ( a_j - a_j^\dag ).
\ee
For an interval with $\ell$ sites on the spin chain in a Gaussian state $\r$, one  can define the $2\ell \times 2\ell$ correlation matrix by
\be \label{Grho}
\lag d_{m_1} d_{m_2} \rag_\r = \d_{m_1m_2} + \G_{m_1m_2}, ~~
m_1,m_2 = 1,2,\cdots,2\ell.
\ee
The $2^\ell \times 2^\ell$ RDM in the state $\r$ is \cite{Vidal:2002rm,Latorre:2003kg}
\be
\r_A = \f{1}{2^\ell} \sum_{s_1,\cdots,s_{2\ell}\in\{0,1\}}
             \lag d_{2\ell}^{s_{2\ell}} \cdots d_1^{s_1} \rag_\r
             d_1^{s_1} \cdots d_{2\ell}^{s_{2\ell}},
\ee
and the multi-point correlation functions $\lag d_{2\ell}^{s_{2\ell}} \cdots d_1^{s_1} \rag_\r$ are calculated from the correlation matrix (\ref{Grho}) by Wick contractions.

For the ground state on an infinite chain $\r(\vn)$, the ground state on a length $L$ circular chain $\r(L)$,  and a thermal state with inverse temperature $\b$ on an infinite chain $\r(\b)$, the nonvanishing components of the correlation matrix $\G$ can be written in terms of the function $g_j$ that is defined as
\be \label{gjdef}
\G_{2j_1-1,2j_2} = - \G_{2j_2,2j_1-1} = g_{j_2-j_1}.
\ee
In the critical Ising chain, we have in different states
\bea \label{confused}
&& g_j(\vn) = - \f{\ii}{\pi} \f{1}{j+\f12}, \nn\\
&& g_j(L) = - \f{\ii}{L} \f{1}{\sin\f{\pi(j+\f12)}{L}}, \nn\\
&& g_j(\b) = - \f{\ii}{\pi} \f{1}{j+\f12}
              +\f{2\ii}{\pi} \int_0^{\pi}\dd\vph \f{\sin[(j+\f12)\vph]}{1+\exp(2\b\sin\f\vph2)}.
\eea
In the XX chain with zero field we obtain
\bea
&& g_j(\vn) = \f{2\ii}{\pi j} \sin \f{\pi j}{2}, ~~ g_0(\vn)=0, \nn\\
&& g_j(L) = \f{2\ii}{L} \f{\sin \f{\pi j}{2}}{\sin\f{\pi j}{L}}, ~~ g_0(L)=0, \nn\\
&& g_j(\b) = \f{2\ii}{\pi j} \sin \f{\pi j}{2}
            -\f{2\ii[1-(-)^j]}{\pi} \int_0^{\f{\pi}{2}}\dd\vph \f{\cos(j\vph)}{1+\exp(\b\cos\vph)}, ~~
   g_0(\b) = 0.
\eea

For a state with both finite size and finite temperature $\r(L,\b)$, it is more complicated to construct the numerical RDM $\r_A(L,\b)$.
Depending on the number of zero modes, i.e. modes with zero energy, we consider three different cases in the following subsections.
In the gapped XY chain there is no zero mode.
In the critical Ising chain and the XX chain with zero field there are respectively one and two zero modes.

\subsection{Gapped XY chain}

There is no zero mode in the gapped XY chain.
The normalized density matrix of the whole system in a thermal state is
\be
\r = \f{\ep^{-\b H}}{\tr\ep^{-\b H}} = \f{\ep^{-\b H_\NS} + \cP \ep^{-\b H_\NS} + \ep^{-\b H_\rR} - \cP\ep^{-\b H_\rR}}{Z_\NS^+ + Z_\NS^-+Z_\rR^+ - Z_\rR^-},
\ee
with
\bea
&& Z_{\rm NS}^+ = \prod_{k \in {\rm NS}} \Big(2\cosh\f{\b\ve_k}{2}\Big),~~
   Z_{\rm NS}^- = \prod_{k \in {\rm NS}} \Big(2\sinh\f{\b\ve_k}{2}\Big), \nn\\
&& Z_{\rm R}^+ = \prod_{k \in {\rm R}} \Big(2\cosh\f{\b\ve_k}{2}\Big), ~~
   Z_{\rm R}^- = \prod_{k \in {\rm R}} \Big(2\sinh\f{\b\ve_k}{2}\Big).
\eea

We can rewrite the thermal density matrix as
\bea
&& \r = \f1{Z_\NS^+ + Z_\NS^-+Z_\rR^+ - Z_\rR^-}
        \Big( Z_\NS^+ \r_\NS^+ + Z_\NS^- \r_\NS^- + Z_\rR^+ \r_\rR^+ - Z_\rR^- \r_\rR^- \Big), \nn\\
&& \r_\NS^+ = \f{\ep^{-\b H_\NS}}{Z_\NS^+}, ~~
   \r_\NS^- = \f{\cP\ep^{-\b H_\NS}}{Z_\NS^-}, ~~
   \r_\rR^+ = \f{\ep^{-\b H_\rR}}{Z_\rR^+}, ~~
   \r_\rR^- = \f{\cP\ep^{-\b H_\rR}}{Z_\rR^-}.
\eea
Note that all the four density matrices $\r_\NS^+$, $\r_\NS^-$, $\r_\rR^+$, $\r_\rR^-$ are Gaussian and properly normalized, and so we can construct their RDMs $\r_{A,\NS}^+$, $\r_{A,\NS}^-$, $\r_{A,\rR}^+$, $\r_{A,\rR}^-$ from the corresponding correlation matrices.
Then we get the RDM of the thermal density matrix
\be
\r_A = \f1{Z_\NS^+ + Z_\NS^-+Z_\rR^+ - Z_\rR^-}
       \Big( Z_\NS^+ \r_{A,\NS}^+ + Z_\NS^- \r_{A,\NS}^- + Z_\rR^+ \r_{A,\rR}^+ - Z_\rR^- \r_{A,\rR}^- \Big).
\ee
For $\r_{A,\NS}^+$, $\r_{A,\NS}^-$, $\r_{A,\rR}^+$, $\r_{A,\rR}^-$, we have the correlation matrix with nonvanishing components (\ref{gjdef}) and
\bea
&& g_j = -\f{\ii}{L}\sum_{k\in\NS}\ep^{\ii(j\vph_k-\th_k)}\tanh\f{\b\ve_k}{2}, \nn\\
&& g_j = -\f{\ii}{L}\sum_{k\in\NS}\ep^{\ii(j\vph_k-\th_k)}\coth\f{\b\ve_k}{2}, \nn\\
&& g_j = -\f{\ii}{L}\sum_{k\in\rR}\ep^{\ii(j\vph_k-\th_k)}\tanh\f{\b\ve_k}{2}, \nn\\
&& g_j = -\f{\ii}{L}\sum_{k\in\rR}\ep^{\ii(j\vph_k-\th_k)}\coth\f{\b\ve_k}{2}.
\eea

\subsection{Critical Ising chain}

There is one zero mode in the R sector, i.e. $\ve_0=0$,  which needs a careful treatment.
We write the thermal density matrix as
\bea
&& \r = \f1{Z_\NS^+ + Z_\NS^-+Z_\rR^+} \Big( Z_\NS^+ \r_\NS^+ + Z_\NS^- \r_\NS^- + Z_\rR^+ \r_\rR^+ - \f{2\td Z_\rR^-\s_1^z}{L} \td\r_\rR^- \Big), \nn\\
&& \r_\NS^+ = \f{\ep^{-\b H_\NS}}{Z_\NS^+}, ~~
   \r_\NS^- = \f{\cP\ep^{-\b H_\NS}}{Z_\NS^-}, ~~
   \r_\rR^+ = \f{\ep^{-\b H_\rR}}{Z_\rR^+}, ~~
   \td\r_\rR^- = \f{\s_1^z\cP\ep^{-\b H_\rR}}{{2\td Z_\rR^-}/{L}}.
\eea
We have defined
\be
\td Z_{\rm R}^- = \prod_{k \in {\rm R},k\neq0} \Big(2\sinh\f{\b\ve_k}{2}\Big).
\ee
Note that the zero mode makes $Z_\rR^-=0$.
We have also defined $\td\r_\rR^-$ following the appendix D of \cite{Fagotti:2010yr}.
The RDM for the thermal density matrix is
\be
\r_A = \f1{Z_\NS^+ + Z_\NS^-+Z_\rR^+} \Big( Z_\NS^+ \r_{A,\NS}^+ + Z_\NS^- \r_{A,\NS}^- + Z_\rR^+ \r_{A,\rR}^+ - \f{2\td Z_\rR^-\s_1^z}{L} \td\r_{A,\rR}^- \Big).
\ee
All the RDMs $\r_{A,\NS}^+$, $\r_{A,\NS}^-$, $\r_{A,\rR}^+$, $\td\r_{A,\rR}^-$ are Gaussian
and the RDMs $\r_{A,\NS}^+$, $\r_{A,\NS}^-$, $\r_{A,\rR}^+$ can be constructed in the same way as that in the previous subsection.
For $\td\r_{A,\rR}^-$, we have the correlation matrix with components
\bea
&& \G_{2j_1-1,2j_2-1} = - \G_{2j_2-1,2j_1-1} = \G_{2j_1,2j_2} = - \G_{2j_2,2j_1} = f_{j_1j_2}, \nn\\
&& \G_{2j_1-1,2j_2} = - \G_{2j_2,2j_1-1} = g_{j_1j_2},
\eea
and definitions
\bea
&& f_{j_1j_2} = - \d_{j_11}+\d_{j_21}, ~~
   g_{j_1j_2} = \td g_{j_2-j_1} + \td g_0 - \td g_{j_2-1} - \td g_{1-j_1}, \nn\\
&& \td g_j = -\f{\ii}{L}\sum_{k\in\rR,k\neq0}\ep^{\ii(j\vph_k-\th_k)}\coth\f{\b\ve_k}{2}.
\eea

\begin{figure}[htbp]
  \centering
  \includegraphics[height=0.25\textwidth]{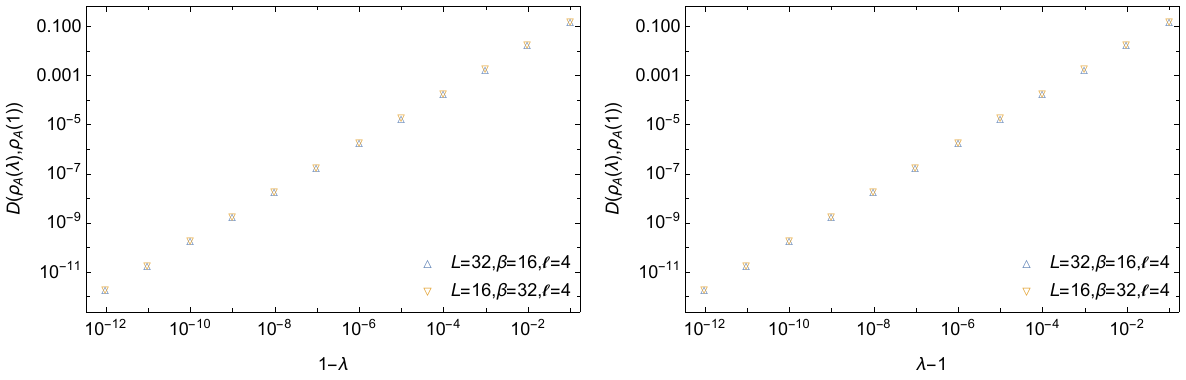}\\
  \caption{Trace distance of the thermal RDM in the gapped Ising chain $\r_A(\l)$ and the thermal RDM in the critical Ising chain $\r_A(1)$.}
  \label{gappedXYIsing}
\end{figure}

To confirm that the above trick works we compare the RDM in the gapped XY chain with $\g=1$ and $\l \to 1$, i.e. gapped Ising chain with $\l \to 1$, which we denote by $\r_A(\l)$, with the RDM in the critical Ising chain, which we denote by $\r_A(1)$.
We plot the trace distance of $\r_A(\l)$ and $\r_A(1)$ in Fig.~\ref{gappedXYIsing}. We see that as $\l \to 1$ the thermal RDM in the gapped Ising chain approaches to the RDM in the critical Ising chain.
By numerical fit, we get approximately
\be
D( \r_A(\l), \r_A(1) ) \pp |\l-1|
\ee
This indicates that the thermal RDM in critical Ising chain we have constructed is correct.

\subsection{XX chain with zero field}

There are two zero modes in the R sector, i.e. $\ve_{\pm L/4}=0$.
Remember that in this paper we only consider the cases that $L$ are four times of integers.
We write the thermal density matrix as
\bea
&& \r = \f1{Z_\NS^+ + Z_\NS^-+Z_\rR^+} \Big( Z_\NS^+ \r_\NS^+ + Z_\NS^- \r_\NS^- + Z_\rR^+ \r_\rR^+ - \f{16\td Z_\rR^-\s_1^z\s_2^z}{L^2} \td\r_\rR^- \Big), \nn\\
&& \r_\NS^+ = \f{\ep^{-\b H_\NS}}{Z_\NS^+}, ~~
   \r_\NS^- = \f{\cP\ep^{-\b H_\NS}}{Z_\NS^-}, ~~
   \r_\rR^+ = \f{\ep^{-\b H_\rR}}{Z_\rR^+}, ~~
   \td\r_\rR^- = \f{\s_1^z\s_2^z\cP\ep^{-\b H_\rR}}{{16\td Z_\rR^-}/{L^2}},
\eea
with the new definition
\be
\td Z_{\rm R}^- = \prod_{k \in {\rm R},k\neq\pm L/4} \Big(2\sinh\f{\b\ve_k}{2}\Big).
\ee
The RDM of the thermal density matrix is
\be
\r_A = \f1{Z_\NS^+ + Z_\NS^-+Z_\rR^+} \Big( Z_\NS^+ \r_{A,\NS}^+ + Z_\NS^- \r_{A,\NS}^- + Z_\rR^+ \r_{A,\rR}^+ - \f{16\td Z_\rR^-\s_1^z\s_2^z}{L^2} \td\r_{A,\rR}^- \Big).
\ee
All the RDMs $\r_{A,\NS}^+$, $\r_{A,\NS}^-$, $\r_{A,\rR}^+$, $\td\r_{A,\rR}^-$ are Gaussian.
The RDMs $\r_{A,\NS}^+$, $\r_{A,\NS}^-$, $\r_{A,\rR}^+$ could be constructed the same as these in the gapped XY chain.
We get $\td\r_{A,\rR}^-$ from the correlation functions
\bea
&& \lag d_{4l_1-3} d_{4l_2-3} \rag = \lag d_{4l_1-1} d_{4l_2-1} \rag
 = \lag d_{4l_1-2} d_{4l_2-2} \rag = \lag d_{4l_1} d_{4l_2} \rag = \d_{j_1j_2}+ (-)^{l_2}\d_{l_11} - (-)^{l_1}\d_{l_21}, \nn\\
&& \lag d_{4l_1-3} d_{4l_2-1} \rag = \lag d_{4l_1-2} d_{4l_2} \rag = 0, \nn\\
&& \lag d_{4l_1-3} d_{4l_2-2} \rag = \lag d_{4l_1-1} d_{4l_2} \rag
 = \td g_{2(l_2-l_1)} + (-)^{l_2} \td g_{2(1-l_1)} + (-)^{l_1} \td g_{2(l_2-1)} + (-)^{l_1+l_2}\td g_{0}, \nn\\
&& \lag d_{4l_1-3} d_{4l_2} \rag
 = \td g_{2(l_2-l_1)+1} + (-)^{l_2} \td g_{3-2l_1} + (-)^{l_1} \td g_{2l_2-1} + (-)^{l_1+l_2}\td g_{1}, \nn\\
&& \lag d_{4l_1-1} d_{4l_2-2} \rag
 = \td g_{2(l_2-l_1)-1} + (-)^{l_2} \td g_{1-2l_1} + (-)^{l_1} \td g_{2l_2-3} + (-)^{l_1+l_2}\td g_{-1},
\eea
with the definition of the function
\be
\td g_j = -\f{\ii}{L}\sum_{k\in\rR,k\neq\pm L/4}\ep^{\ii j\vph_k}\coth\f{\b\ve_k}{2}.
\ee
Note that $\lag d_{m_1} d_{m_2} \rag = 2 \d_{m_1m_2} - \lag d_{m_2} d_{m_1} \rag$.

\begin{figure}[htbp]
  \centering
  \includegraphics[height=0.25\textwidth]{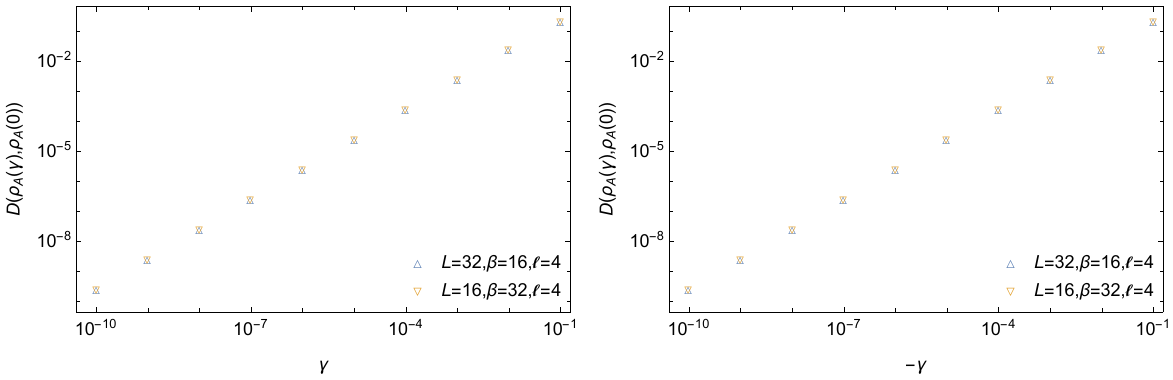}\\
  \caption{Trace distance of the thermal RDM in the gapped XY chain $\r_A(\g)$ and the thermal RDM in the XX chain with no field $\r_A(0)$.}\label{gappedXYXX}
\end{figure}

To confirm that the numerical RDM in the XX chain with zero field is correct we compare it with the RDM in the gapped XY chain with $\l = 0$ and $\g \to 0$, which we denote by $\r_A(\g)$.
We denote the RDM of the XX chain with no field as $\r_A(0)$.
We plot the trace distance of $\r_A(\g)$ and $\r_A(0)$ in Fig.~\ref{gappedXYXX}.
We see that as $\g \to 0$ the thermal RDM in the gapped XY chain approaches to the RDM in the XX chain.
By numerical fit, we get approximately
\be
D( \r_A(\g), \r_A(0) ) \pp |\g|.
\ee

\section{Relative entropy among RDMs in low-lying energy eigenstates}\label{appSRE}

We revisit the relative entropy among the RDMs of one interval $A=[0,\ell]$ on a cylinder with circumference $L$ in various low-lying energy eigenstates, generalizing \cite{Ruggiero:2016khg,Nakagawa:2017fzo,Zhang:2019itb}. With the formula (\ref{PR}), which could be found in \cite{Caputa:2016yzn}, we calculate the relative entropy of an interval with a relatively large length. This checks various results of the exact relative entropy, not only the leading order results in a short interval expansion but also the results with a long interval.

\subsection{Free massless Majorana fermion theory}

\begin{figure}[htbp]
  \centering
  \includegraphics[height=0.35\textwidth]{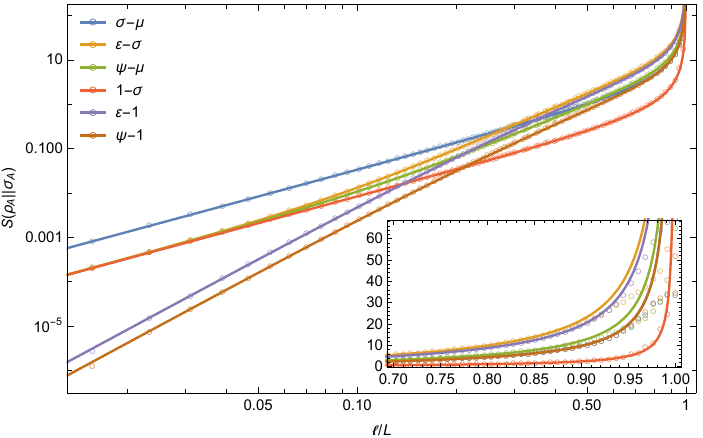}\\
  \caption{Relative entropy of the RDMs in low-lying energy eigenstates in the 2D free massless Majorana fermion theory (solid lines) and critical Ising chain (small empty circles). We have set $L=128$.}\label{IsingRELEE}
\end{figure}

In a 2D CFT, we denote $\r_{A,\cO}=\tr_{\bar A} | \cO\rag \lag \cO|$ as the RDM of $A$ in the excited state $|\cO\rag$ on a cylinder.
In the free massless Majorana fermion theory we consider the primary operators $1,\s,\m,\psi,\bar\psi,\ve$ with conformal weights (0,0), (1/16,1/16), (1/16,1/16), (1/2,0), (0,1/2), (1/2,1/2), respectively.
There are exact results \cite{Ruggiero:2016khg,Nakagawa:2017fzo,Zhang:2019itb} which reads
\bea \label{REeesMajorana}
&& S(\r_{A,1}\|\r_{A,\s}) = S(\r_{A,\s}\|\r_{A,1})
 = S(\r_{A,1}\|\r_{A,\m}) = S(\r_{A,\m}\|\r_{A,1})
 = \f14 \Big( 1- \f{\pi\ell}{L}\cot\f{\pi\ell}{L} \Big), \nn\\
&& S(\r_{A,\s}\|\r_{A,\m}) = S(\r_{A,\m}\|\r_{A,\s}) = 1- \f{\pi\ell}{L}\cot\f{\pi\ell}{L}, \nn\\
&& S(\r_{A,\psi}\|\r_{A,1}) = S(\r_{A,\bar\psi}\|\r_{A,1}) = S(\r_{A,\ve}\|\r_{A,\psi}) = S(\r_{A,\ve}\|\r_{A,\bar\psi}) = 1- \f{\pi\ell}{L}\cot\f{\pi\ell}{L}
                                                 + \sin \f{\pi\ell}{L} \nn\\
&& \phantom{S(\r_{A,\psi}\|\r_{A,1}) = S(\r_{A,\bar\psi}\|\r_{A,1}) = S(\r_{A,\ve}\|\r_{A,\psi}) = S(\r_{A,\ve}\|\r_{A,\bar\psi}) =}
                                                 + \log\Big( 2 \sin \f{\pi\ell}{L} \Big)
                                                 + \psi\Big(\f12 \csc\f{\pi\ell}{L}\Big), \nn
\eea\bea
&& S(\r_{A,\ve}\|\r_{A,1}) = 2 \Big( 1- \f{\pi\ell}{L}\cot\f{\pi\ell}{L} \Big)
                  + 2 \Big[ \sin \f{\pi\ell}{L}
                          + \log\Big( 2 \sin \f{\pi\ell}{L} \Big)
                          + \psi\Big(\f12 \csc\f{\pi\ell}{L}\Big) \Big],  \\
&& S(\r_{A,\psi}\|\r_{A,\s}) = S(\r_{A,\psi}\|\r_{A,\m}) = S(\r_{A,\bar\psi}\|\r_{A,\s}) = S(\r_{A,\bar\psi}\|\r_{A,\m})
 = \f54 \Big( 1- \f{\pi\ell}{L}\cot\f{\pi\ell}{L} \Big)
 + \sin \f{\pi\ell}{L} \nn\\
&& \phantom{S(\r_{A,\psi}\|\r_{A,\s}) = S(\r_{A,\psi}\|\r_{A,\m}) = S(\r_{A,\bar\psi}\|\r_{A,\s}) = S(\r_{A,\bar\psi}\|\r_{A,\m}) = }
 + \log\Big( 2 \sin \f{\pi\ell}{L} \Big)
 + \psi\Big(\f12 \csc\f{\pi\ell}{L}\Big), \nn\\
&& S(\r_{A,\ve}\|\r_{A,\s}) = S(\r_{A,\ve}\|\r_{A,\m}) = \f{9}{4} \Big( 1- \f{\pi\ell}{L}\cot\f{\pi\ell}{L} \Big)
                    + 2 \Big[ \sin \f{\pi\ell}{L}
                            + \log\Big( 2 \sin \f{\pi\ell}{L} \Big)
                            + \psi\Big(\f12 \csc\f{\pi\ell}{L}\Big) \Big]. \nn
\eea
We compare some of the analytical CFT results with the numerical spin chain results in Fig.~\ref{IsingRELEE}.
Generally, we see good matches not only for a short interval, but also for a long interval.
Specially, the relative entropies $S(\r_{A,\ve}\|\r_{A,\s})$, $S(\r_{A,\psi}\|\r_{A,\m})$, $S(\r_{A,1}\|\r_{A,\s})$ have the same leading order short interval expansion results, but they are different for a long interval, as we can see in both the CFT and the spin chain results in the figure.
In some cases there are mismatches as $\ell/L \to 1$, and we attribute them to numerical errors in the spin chain calculations.
Actually, in the limit $\ell/L \to 1$ all the relative entropies (\ref{REeesMajorana}) in CFT are divergent, as they approach relative entropies of two pure states.

\subsection{Free massless Dirac fermion theory}

\begin{figure}[htbp]
  \centering
  \includegraphics[height=0.35\textwidth]{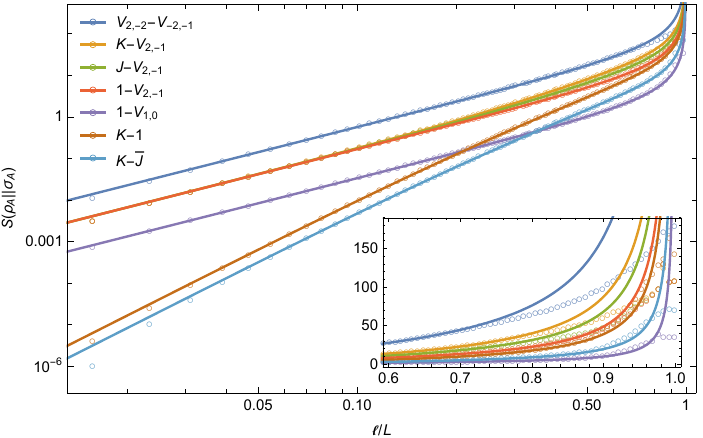}\\
  \caption{Relative entropy of the RDMs in low-lying energy eigenstates in the 2D free massless Dirac fermion theory (solid lines) and the XX chain with zero field (small empty circles). We have set $L=128$.}\label{XXRELEE}
\end{figure}

For the 2D free massless Dirac fermion theory, it is convenient to use the language of the 2D free massless compact boson theory with the unit target space radius.
We consider the RDMs in the excited states by the primary operators 1, $V_{\a,\bar\a}$, $J$, $\bar J$, $K=J\bar J$ with conformal weights (0,0), $(\a^2/2,\bar\a^2/2)$, (1,0), (0,1), (1,1), respectively.
There are the following exact results \cite{Lashkari:2014yva,Lashkari:2015dia,Ruggiero:2016khg,Nakagawa:2017fzo,Zhang:2019itb}
\bea
&& S(\r_{A,V_{\a,\bar\a}}\|\r_{A,V_{\a',\bar\a'}}) = [(\a-\a')^2+(\bar\a-\bar\a')^2]\Big( 1- \f{\pi\ell}{L}\cot\f{\pi\ell}{L} \Big), \nn\\
&& S(\r_{A,J}\|\r_{A,V_{\a,\bar\a}}) = S(\r_{A,\bar J}\|\r_{A,V_{\a,\bar\a}})
                           =  (2 + \a^2+\bar\a^2)\Big( 1- \f{\pi\ell}{L}\cot\f{\pi\ell}{L} \Big) \nn \\
&& \phantom{S(\r_{A,J}\|\r_{A,V_{\a,\bar\a}}) = S(\r_{A,\bar J}\|\r_{A,V_{\a,\bar\a}})=}
                            + 2 \Big[ \sin \f{\pi\ell}{L}
                            + \log\Big( 2 \sin \f{\pi\ell}{L} \Big)
                            + \psi\Big(\f12 \csc\f{\pi\ell}{L}\Big) \Big], \\
&& S(\r_{A,K}\|\r_{A,V_{\a,\bar\a}}) = (4 + \a^2+\bar\a^2)\Big( 1- \f{\pi\ell}{L}\cot\f{\pi\ell}{L} \Big)
                            + 4 \Big[ \sin \f{\pi\ell}{L}
                            + \log\Big( 2 \sin \f{\pi\ell}{L} \Big)
                            + \psi\Big(\f12 \csc\f{\pi\ell}{L}\Big) \Big], \nn\\
&& S(\r_{A,K}\|\r_{A,J}) = S(\r_{A,K}\|\r_{A,\bar J})
                           =  2 \Big( 1- \f{\pi\ell}{L}\cot\f{\pi\ell}{L} \Big)
                            + 2 \Big[ \sin \f{\pi\ell}{L}
                            + \log\Big( 2 \sin \f{\pi\ell}{L} \Big)
                            + \psi\Big(\f12 \csc\f{\pi\ell}{L}\Big) \Big]. \nn
\eea
We compare the some of the analytical CFT results with the numerical CFT results in Fig.~\ref{XXRELEE}.

\providecommand{\href}[2]{#2}\begingroup\raggedright\endgroup


\end{document}